%% file: outflows_masses.tex
\documentclass[twocolumn]{aastex63}

\usepackage{amsmath,amssymb}
\usepackage{xspace}
\usepackage{CJK}

\usepackage{hyperref}
\hypersetup{
    colorlinks=true,
    linkcolor=blue,
    filecolor=magenta,      
    urlcolor=cyan,
}

\usepackage[normalem]{ulem} 
\usepackage{cancel} 

\usepackage{color}

\newcommand{\um}{\ensuremath{\mu\rm{m}}\xspace}
\newcommand{\kms}{\ensuremath{\rm{km\,s}^{-1}}\xspace}
\newcommand{\alphaco}{\ensuremath{\alpha_{\rm{CO}}}\xspace}

\newcommand{\sigsfr}{\ensuremath{\Sigma_{\rm{SFR}}}\xspace}
\newcommand{\sigir}{\ensuremath{\Sigma_{\rm{IR}}}\xspace}

\newcommand{\lfir}{\ensuremath{L_{\rm{FIR}}}\xspace}
\newcommand{\lir}{\ensuremath{L_{\rm{IR}}}\xspace}
\newcommand{\fagn}{\ensuremath{f_\mathrm{AGN}}\xspace}

\newcommand{\Tdust}{\ensuremath{T_{\rm{dust}}}\xspace}

\newcommand{\rdust}{\ensuremath{r_{\rm{dust}}}\xspace}

\newcommand{\MHt}{\ensuremath{M_{\rm{H}_2}}\xspace}

\newcommand{\Msol}{\ensuremath{\rm{M}_\odot}\xspace}
\newcommand{\Lsol}{\ensuremath{\rm{L}_\odot}\xspace}
\newcommand{\LAGN}{\ensuremath{L_{\rm{AGN}}}\xspace}
\newcommand{\LSF}{\ensuremath{L_{\rm{SF}}}\xspace}

\newcommand{\tdepsf}{\ensuremath{t_{\rm{dep,\,SF}}}\xspace}
\newcommand{\tdepout}{\ensuremath{t_{\rm{dep,\,out}}}\xspace}

\newcommand{\vmax}{\ensuremath{v_\mathrm{max}}\xspace}
\newcommand{\vfifty}{\ensuremath{v_{50}}\xspace}
\newcommand{\vef}{\ensuremath{v_{84}}\xspace}
\newcommand{\cii}{[C{\scriptsize II}]\xspace}

\newcommand{\Mout}{\ensuremath{M_{\mathrm{out}}}\xspace}
\newcommand{\Rout}{\ensuremath{R_{\mathrm{out}}}\xspace}
\newcommand{\vout}{\ensuremath{v_{\mathrm{out}}}\xspace}

\newcommand{\etaout}{\ensuremath{\eta_{\mathrm{out}}}\xspace}
\newcommand{\Mdot}{\ensuremath{\dot{M}_{\mathrm{out}}}\xspace}
\newcommand{\pdot}{\ensuremath{\dot{p}_{\mathrm{out}}}\xspace}
\newcommand{\Edot}{\ensuremath{\dot{E}_{\mathrm{out}}}\xspace}
\newcommand{\fcov}{\ensuremath{f_{\mathrm{cov}}}\xspace}
\newcommand{\EWvth}{\ensuremath{\mathrm{EW}_{v < -200}}\xspace}
\newcommand{\EWvoh}{\ensuremath{\mathrm{EW}_{v < -100}}\xspace}
\newcommand{\EWtot}{\ensuremath{\mathrm{EW}_{\mathrm{total}}}\xspace}

\shortauthors{J.~S.~Spilker, et~al.}
\shorttitle{Ubiquitous $z$$>$4 Molecular Outflows II. Outflow Properties}

\begin{document}
\begin{CJK*}{UTF8}{gbsn}

\defcitealias{spilker18a}{S18}
\defcitealias{spilker20}{Paper~I}
\defcitealias{spilker20a}{Paper~II}
\defcitealias{gonzalezalfonso17}{GA17}
\defcitealias{herreracamus20}{HC20}

\title{Ubiquitous Molecular Outflows in $z$ $>$ 4 Massive, Dusty Galaxies\\
II. Momentum-Driven Winds Powered by Star Formation in the Early Universe}

\correspondingauthor{Justin S. Spilker}
\email{spilkerj@gmail.com}

\author[0000-0003-3256-5615]{Justin S. Spilker}
\altaffiliation{NHFP Hubble Fellow}
\affiliation{Department of Astronomy, University of Texas at Austin, 2515 Speedway, Stop C1400, Austin, TX 78712, USA}

\author[0000-0002-6290-3198]{Manuel~Aravena}
\affiliation{N\'{u}cleo de Astronom{\'i}a de la Facultad de Ingenier\'{i}a y Ciencias, Universidad Diego Portales, Av. Ej\'{e}rcito Libertador 441, Santiago, Chile}

\author[0000-0001-7946-557X]{Kedar~A.~Phadke}
\affiliation{Department of Astronomy, University of Illinois, 1002 West Green St., Urbana, IL 61801, USA}

\author[0000-0002-3915-2015]{Matthieu~B{\'e}thermin}
\affiliation{Aix Marseille Univ., CNRS, CNES, LAM, Marseille, France}

\author{Scott~C.~Chapman}
\affiliation{Department of Physics and Astronomy, University of British Columbia, 6225 Agricultural Rd., Vancouver, V6T 1Z1, Canada}
\affiliation{National Research Council, Herzberg Astronomy and Astrophysics, 5071 West Saanich Rd., Victoria, V9E 2E7, Canada}
\affiliation{Department of Physics and Atmospheric Science, Dalhousie University, Halifax, Nova Scotia, Canada}

\author[0000-0002-5823-0349]{Chenxing~Dong~(董辰兴)}
\affiliation{Department of Astronomy, University of Florida, 211 Bryant Space Sciences Center, Gainesville, FL 32611, USA}

\author[0000-0002-0933-8601]{Anthony~H.~Gonzalez}
\affiliation{Department of Astronomy, University of Florida, 211 Bryant Space Sciences Center, Gainesville, FL 32611, USA}

\author[0000-0003-4073-3236]{Christopher~C.~Hayward}
\affiliation{Center for Computational Astrophysics, Flatiron Institute, 162 Fifth Avenue, New York, NY, 10010, USA}

\author[0000-0002-8669-5733]{Yashar~D.~Hezaveh}
\affiliation{D\'{e}partement de Physique, Universit\'{e} de Montr\'{e}al, Montreal, Quebec, H3T 1J4, Canada}
\affiliation{Center for Computational Astrophysics, Flatiron Institute, 162 Fifth Avenue, New York, NY, 10010, USA}

\author[0000-0002-4208-3532]{Katrina~C.~Litke}
\affiliation{Steward Observatory, University of Arizona, 933 North Cherry Avenue, Tucson, AZ 85721, USA}

\author[0000-0001-6919-1237]{Matthew~A.~Malkan}
\affiliation{Department of Physics and Astronomy, University of California, Los Angeles, CA 90095-1547, USA}

\author[0000-0002-2367-1080]{Daniel~P.~Marrone}
\affiliation{Steward Observatory, University of Arizona, 933 North Cherry Avenue, Tucson, AZ 85721, USA}

\author[0000-0002-7064-4309]{Desika~Narayanan}
\affiliation{Department of Astronomy, University of Florida, 211 Bryant Space Sciences Center, Gainesville, FL 32611, USA}
\affiliation{University of Florida Informatics Institute, 432 Newell Drive, CISE Bldg E251, Gainesville, FL 32611, USA}
\affiliation{Cosmic Dawn Center (DAWN), DTU-Space, Technical University of Denmark, Elektrovej 327, DK-2800 Kgs. Lyngby, Denmark}

\author[0000-0001-7477-1586]{Cassie~Reuter}
\affiliation{Department of Astronomy, University of Illinois, 1002 West Green St., Urbana, IL 61801, USA}

\author[0000-0001-7192-3871]{Joaquin~D.~Vieira}
\affiliation{Department of Astronomy, University of Illinois, 1002 West Green St., Urbana, IL 61801, USA}
\affiliation{Department of Physics, University of Illinois, 1110 West Green St., Urbana, IL 61801, USA}
\affiliation{National Center for Supercomputing Applications, University of Illinois, 1205 West Clark St., Urbana, IL 61801, USA}

\author[0000-0003-4678-3939]{Axel~Wei{\ss}}
\affiliation{Max-Planck-Institut f\"{u}r Radioastronomie, Auf dem H\"{u}gel 69, D-53121 Bonn, Germany}

\begin{abstract}

Galactic outflows of molecular gas are a common occurrence in galaxies and may represent a mechanism by which galaxies self-regulate their growth, redistributing gas that could otherwise have formed stars. We previously presented the first survey of molecular outflows at $z$$>$4 towards a sample of massive, dusty galaxies. Here we characterize the physical properties of the molecular outflows discovered in our survey. Using low-redshift outflows as a training set, we find agreement at the factor-of-two level between several outflow rate estimates. We find molecular outflow rates 150--800\,\Msol/yr and infer mass loading factors just below unity. Among the high-redshift sources, the molecular mass loading factor shows no strong correlations with any other measured quantity. The outflow energetics are consistent with expectations for momentum-driven winds with star formation as the driving source, with no need for energy-conserving phases. There is no evidence for AGN activity in our sample, and while we cannot rule out deeply-buried AGN, their presence is not required to explain the outflow energetics, in contrast to nearby obscured galaxies with fast outflows. The fraction of the outflowing gas that will escape into the circumgalactic medium (CGM), though highly uncertain, may be as high as 50\%. This nevertheless constitutes only a small fraction of the total cool CGM mass based on a comparison to $z$$\sim$2--3 quasar absorption line studies, but could represent $\gtrsim$10\% of the CGM metal mass. Our survey offers the first statistical characterization of molecular outflow properties in the very early universe.

\end{abstract}

\section{Introduction} \label{intro}

Powerful galactic outflows or winds have been widely invoked in the establishment and regulation of many fundamental observed correlations in galaxies. Outflows driven by supermassive black hole feedback or processes related to star formation (e.g. stellar winds, supernovae, radiation pressure) are thought to regulate the growth of both the black hole and the stellar component of galaxies \citep[e.g.][]{silk98,fabian99,gebhardt00}. Outflows are also invoked as an important mechanism regulating the metallicity of galaxies, capable of transporting heavy elements into the circumgalactic medium that surrounds galaxies \citep[e.g.][]{tumlinson17}. They are also likely necessary to explain the rapid suppression (`quenching') of star formation in massive galaxies and the resulting global and spatially-resolved properties of the stars and gas in quenched galaxies at high redshift \citep[e.g.][]{tacchella15,barro16,spilker18b,bezanson19,spilker19}.

Feedback and outflows are widely viewed as necessary in simulations in order to prevent over-cooling and consequently overly-massive galaxies. Feedback is typically included in ad hoc ways, and prescriptions differ greatly across simulations (see \citealt{somerville15} for a recent review). Recent high-resolution zoom simulations have been able to drive outflows self-consistently \citep[e.g.][]{muratov15,agertz16}, but the galaxy parameter space probed is still limited (usually focusing on Milky Way-like halos). Thus, constraining outflow scaling relations is useful both for testing predictions from high-resolution simulations and for informing sub-grid prescriptions used in large-volume simulations.

Outflows appear to be ubiquitous in galaxies, and the winds are known to span many orders of magnitude in temperature and density \citep[e.g.][]{thompson16,schneider17}, and as such various components of the winds are observable from X-ray to radio wavelengths \citep[e.g.][]{leroy15}. The cold molecular component of outflows is of special interest for many reasons,  not least of which is that molecular gas is the raw fuel for future star formation and appears to be the largest component by mass of most outflows (see \citealt{veilleux20} for a recent review). The cold gas in outflows is notoriously difficult to reproduce in simulations because the thermal balance of outflowing gas depends on the detailed hydrodynamics and heating/cooling processes on spatial scales much smaller than typically achieved \citep[e.g.][]{scannapieco13,schneider17}. The molecular outflow properties of large samples of galaxies can thus provide a valuable constraint for cosmological galaxy formation simulations \citep[e.g.][]{muratov15,dave19,hayward20}.

In the first paper in this series (\citealt{spilker20}, hereafter \citetalias{spilker20}), we presented the first sample of molecular outflows in the $z>4$ universe, using Atacama Large Millimeter Array (ALMA) observations of the hydroxyl (OH) 119\,\um doublet as an outflow tracer. The sample was selected from the South Pole Telescope (SPT) sample of gravitationally lensed dusty star-forming galaxies (DSFGs), targeting intrinsically luminous galaxies, $\log \lir/\Lsol \sim 12.5-13.5$. We found unambiguous outflows in 8/11 ($\sim$75\%) of the sample, approximately tripling the number of known molecular winds at $z>4$. The observations also spatially resolved the outflows, and we found evidence for clumpy substructure in the outflows on scales of $\sim$500\,pc. 

High-redshift DSFGs such as those targeted by our sample, in particular, can offer unique insight into the physics of feedback and its role in galaxy evolution. Their star formation rates (SFRs) and SFR surface densities are unprecedented in the local universe, and approach the theoretical maximum momentum injection rate from stellar feedback beyond which the remaining gas is unbound \citep[e.g.][]{murray05,thompson05}. DSFGs are expected to trace the most massive dark matter halos of their epoch, offering insight into galaxy formation in dense environments \citep[e.g.][]{marrone18,miller18,long20}. They are also one of few viable populations capable of producing massive quiescent galaxies now identified at $z\sim4$, the existence of which implies that the progenitor systems must have experienced powerful and effective feedback in order to suppress star formation \citep[e.g.][]{straatman14,toft14}.

Our primary focus in this work is to understand the physical properties of the molecular outflows we have detected at $z>4$ based on the measured OH 119\,\um profiles. This is made difficult by the fact that the 119\,\um line opacity is expected to be very high, $\tau_{\mathrm{OH}\,119\um} \gtrsim 10$ \citep{fischer10}. In the nearby universe, extensive observations of ULIRGs and obscured QSOs with \textit{Herschel}/PACS allowed many OH lines to be detected towards the same objects, including some transitions with far lower optical depths and excited transitions that can only arise from the warmest and densest regions. With many OH transitions, self-consistent radiative transfer modeling can reproduce all observed line profiles as well as the dust continuum emission simultaneously (\citealt{gonzalezalfonso17}; hereafter \citetalias{gonzalezalfonso17}). In the distant universe we are unlikely to possess such a rich trove of information until the next generation of far-IR space missions become reality. Even with ALMA at the highest redshifts the atmosphere precludes observations of the full suite of OH diagnostics, and the other OH transitions are typically weaker than the 119\,\um ground state lines. It thus behooves us to understand whether and how well outflow properties can be determined if only the 119\,\um OH doublet has been detected. We presented a first attempt at such an analysis in \citet[][hereafter \citetalias{spilker18a}]{spilker18a}, which we expand upon here.

Readers interested only in our interpretation of the outflow properties we derive here are welcome to skip to Section~\ref{results}, which presents our main findings and discussion. Section~\ref{outflowcalcs} gives an overview of our assumed outflow geometry, calculations of outflow properties, and the literature reference samples we use as a training set for our own sample galaxies. Section~\ref{outflowratemethods} describes the different methods we use to estimate the outflow rates and explores the level of agreement between the methods. We summarize and conclude in Section~\ref{conclusions}. We assume a flat $\Lambda$CDM cosmology with $\Omega_m=0.307$ and $H_0=67.7$\,\kms\,Mpc$^{-1}$ \citep{planck15}, and we take the total infrared and far-infrared luminosities \lir and \lfir to be integrated over rest-frame 8--1000 and 40--120\,\um, respectively. We assume a conversion between \lir and SFR of SFR$= 1.49 \times 10^{-10} \lir$, with \lir in \Lsol and SFR in \Msol/yr \citep{murphy11}. Tables of the outflow properties from this work, as well as the SPT sample properties from \citetalias{spilker20}, are available in electronic form at \url{https://github.com/spt-smg/publicdata}.

\section{Outflow Assumptions and Literature Reference Sample} \label{outflowcalcs}

\subsection{Assumed Outflow Geometry} \label{outflowgeom}

Where necessary throughout this work, we assume a spherical `time-averaged thin shell' geometry widely used in the literature \citep[see][]{rupke05b}, in which a mass-conserving outflow with constant outflow rate expands following a density profile $n \propto r^{-2}$. In this geometry, the outflow rate and mass are related through
\begin{subequations}\label{eq:mdot}
\begin{equation}
  \Mdot = 4 \pi \Rout^2 \: \overline{\mu} \, m_H \: N_H \: \vout / \Rout
\end{equation}
\begin{equation}
  \Mout = \Mdot \Rout / \vout
\end{equation}
\end{subequations}
with \vout the characteristic outflow velocity, \Rout the outflow inner radius, $\overline{\mu} = 1.4$ the mean mass per hydrogen atom (including the cosmological helium abundance) and $m_H$ the mass of a hydrogen atom. These quantities are fundamentally linked to the column density of gas along the line of sight $N_H$ responsible for generating the observed absorption profiles. If we drop the assumption of spherical symmetry, these quantities are also linearly proportional to the covering fraction \fcov, the fraction of the full $4\pi$\,sr covered by wind material as seen from the source. For our sample, in which the outflows are detected in absorption, the inferred energetics obviously depend strongly on the orientation of the outflow since no absorption can be detected for outflowing material that does not intersect the line of sight to the galaxy. Redshifted receding material is also difficult to constrain for our high-redshift sample.  The ALMA bandwidth probes only a limited range of redshifted velocities, and it is possible for the galaxy itself to be optically thick to emission from the receding material even if the spectral coverage were extended to more redshifted velocities.

As has been discussed extensively in the literature, this assumed geometry leads to more conservative outflow energetics than other simple geometries \citep[e.g.][and references therein]{veilleux20}. In particular the outflow rates are a factor of 3 lower than if the outflow volume is filled with uniform density (which implies a decreasing outflow rate over time for constant flow velocity), and a factor $\Delta r / \Rout$ lower than the `local' or `instantaneous' rate if the wind arises from a thin shell of width $\Delta r$. We take the characteristic velocity to be \vef, the velocity above which 84\% of the absorption occurs. This is also fairly conservative: clearly the maximum velocity is not a `characteristic' outflow velocity, but \vef is more robust to uncertainties in the systemic redshift than the median absorption velocity \vfifty, given the deep absorption at systemic velocities present in most of our sources \citepalias{spilker20}. Finally, we take \Rout to be \rdust, the effective radius of the dust emission at rest-frame $\approx100$\,\um. This radius has been directly measured for the low-redshift literature reference sources by \textit{Herschel}/PACS \citep{lutz16} and from our lensing reconstructions of the ALMA OH continuum data, and is motivated by our observation that the OH outflow absorption is frequently strongest in equivalent width not in the nuclear regions but towards the outskirts \citepalias{spilker20}.

The momentum and kinetic energy outflow rates are then given by
\begin{equation}\label{eq:momen}
  \pdot = \Mdot \vout,\; \Edot = \frac{1}{2}\Mdot \vout^2,
\end{equation}
where the expression for kinetic power assumes negligible contribution from turbulent (i.e. non-bulk) sources.

Although maps of the molecular outflows at $\sim$500\,pc resolution are available for our SPT sample due to their gravitationally lensed nature \citepalias{spilker20}, we do not attempt to match, or otherwise account for, the structures seen in these maps when estimating outflow rates.

\subsection{OH Outflow Training Sample} \label{litref}

In order to determine whether and how well the outflow properties for our high-redshift sample can be estimated given the sole available OH 119\,\um transitions, we compare extensively to low-redshift ULIRGs and obscured QSOs with rich OH data and radiative transfer models from \textit{Herschel}/PACS. In particular, \citetalias{gonzalezalfonso17} self-consistently model 12 nearby IR-luminous galaxies with detections of the OH transitions at 119, 84, 79, and 65\,\um, all of which showed either P Cygni profiles or blueshifted line wings in the 119\,\um doublet. The 84 and 65\,\um doublets are highly-excited lines with lower levels 120 and 300\,K above the ground state that require an intense and warm IR radiation field to be detected, and the cross-ladder 79\,\um doublet has an optical depth $\approx$40$\times$ lower than the 119\,\um transition. We supplement this sample with one additional ULIRG with OH radiative transfer modeling \citep{tombesi15,veilleux17}, and an additional four sources with outflow rates based on the detection of high-velocity CO line wings that were also observed in OH 119\,\um. These four sources have lower typical \lir and lower outflow rates than the primary OH-based sample. We consider these final sources because \citet{lutz20} find reasonable agreement between CO-based and OH-based outflow properties, but we exclude them from our later empirical outflow rate estimation out of an abundance of caution. These sources can at some level be considered akin to a small cross-validation sample, although the dynamic range in outflow properties they span is small.

For all sources, we remeasured various properties of the OH 119\,\um spectra in the same way as for our high-redshift sample \citepalias{spilker20}, including the broadening from the FWHM$\approx$300\,\kms PACS instrumental spectral resolution at these wavelengths. As for our sample, we use the fits to the spectra to measure the velocities above which 50 and 84\% of the absorption takes place, \vfifty and \vef, and the `maximum' outflow velocity \vmax that we take to be the velocity above which 98\% of the absorption takes place. We also measure the total equivalent widths of the absorption components as well as the equivalent widths integrated over various blueshifted velocity ranges; for example, \EWvth refers to the equivalent width integrated over velocities more blueshifted than $-$200\,\kms. We note that all these quantities are non-parametric and therefore depend little on the exact methods used to fit the PACS spectra.

\section{Outflow Rate Estimates} \label{outflowratemethods}

In this section we detail a number of different methods we use to estimate the outflow rates for our $z>4$ SPT DSFG sample. For each method presented here, we estimate uncertainties on the derived outflow rates through a Monte Carlo procedure, repeatedly resampling the measurements within the uncertainties, redoing the fitting analysis, and remeasuring the predicted outflow rates based on the results of each fit. While we provide several empirical fitting formulae that can be used to estimate outflow rates from OH 119\,\um data, we caution that the broad applicability of these formulae is questionable. The literature reference sources are not broadly representative of star-forming galaxies (nor is our $z>4$ sample), consisting solely of IR-luminous galaxies. It is unclear if the conversions we find here can (or should) be extrapolated to less-extreme sources.

For each of our methods here, we aim to find correlations between the published outflow rates for our training sample and the measured OH and ancillary galaxy properties, as detailed in the following subsections. For our objects, we provide observational details and sample properties in \citetalias{spilker20}, and briefly reprise here. Gravitational lensing magnification factors were measured from lens models of the rest-frame 119\,\um dust continuum emission observed by ALMA along with the OH spectroscopy. From simple fits to the OH spectra, we measured basic observed properties such as equivalent widths and velocities. Molecular gas masses were estimated from CO(2--1) detections of all sources, following \citet{aravena16}. Total IR luminosities were measured by fitting to the well-sampled far-IR/submm photometry, which spans rest-frame $\approx$15--600\,\um for all sources. We constrain the contribution of AGN to the total luminosity using rest-frame mid-IR $\sim$15--30\,\um photometry from \textit{Herschel}/PACS sensitive to hot AGN-heated dust near the torus. No source shows evidence of an AGN in the mid-IR (or in any other data, e.g. \citealt{ma16}), with fractional contributions to the total luminosity $\fagn \lesssim 0.1-0.45$ (1$\sigma$; mean upper limit $\fagn \lesssim 0.25$), depending on the source. It is possible that this method underestimates \fagn for heavily-obscured AGN, but we do not know whether such AGN are present in our sample or how common they are if so. In our subsequent analysis, we detail changes to our interpretation that would result from a factor-of-2 underestimate of \fagn (and consequent decrease in the fraction of \lir arising from star formation).

While our parent sample consists of 11 $z>4$ DSFGs, all of which were detected in OH 119\,\um absorption, we determined in \citetalias{spilker20} that only 8 of these show unambiguous evidence for outflows. It is essentially not possible to set upper limits on the outflow properties for the remaining 3 sources, since this would require prior knowledge of, for example, the outflow velocities. Lack of sensitivity is not the issue; all 3 were detected in OH absorption, but ancillary spectral information from \cii or CO data made the OH profiles difficult to interpret conclusively as evidence for outflows. The sources we selected for OH observations are not obviously biased with respect to the full sample of $z>4$ SPT DSFGs in terms of \lir, dust mass, or effective dust temperature \citep{reuter20}, although they are by no means representative of `typical' galaxies at this epoch.

The outflow rates, masses, and energetics we derive for our high-redshift sources from all methods are given in Table~\ref{tab:sptmdots}. These values, as well as the SPT DSFG observed properties given in \citetalias{spilker20} we use to derive the outflow rates, are available in machine-readable format at \url{https://github.com/spt-smg/publicdata}.

\input{table1_outflowrates.tex}

\subsection{Simple Optically-Thin Model} \label{optthin}

We first consider a simple analytic calculation of the outflow rates for our sources and the literature reference sample assuming the OH 119\,\um absorption is optically thin. As already discussed, we expect this to be a very bad assumption, but this calculation does at least provide a hard lower bound on the true outflow rate and an opportunity to determine if some overall correction factor to the optically thin outflow rates could allow a more realistic estimate. While for many nearby galaxies other OH transitions with far lower line opacities can be observed (e.g. the 79\,\um doublet, or lines of the less-abundant $^{18}$OH isotopologue), we must instead attempt to find some other quantity that can provide an empirical correction.

Under the assumption that the absorption is optically thin, the minimum column density of OH molecules $N_{\mathrm{OH}}$ is given by
\begin{equation} \label{eq:NOH}
N_\mathrm{OH} = \frac{ 8 \pi Q_\mathrm{rot}(T_\mathrm{ex}) }{\lambda^3 g_u A_{ul}} 
                \frac{ \mathrm{exp}(E_l / T_\mathrm{ex}) }
                     {1 - \mathrm{exp}(\frac{-h \nu}{k_B T_\mathrm{ex}})}
                \int \tau(v) dv,
\end{equation}
where $\lambda$ and $\nu$ are the wavelength and frequency of the transition, $h$ and $k_B$ are the Planck and Boltzmann constants, $A_{ul}$ is the Einstein `A' coefficient of the transition, $g_u$ the degeneracy of the upper energy level, $E_l$ the lower energy level in temperature units, $Q_\mathrm{rot}$ the rotational partition function evaluated at excitation temperature $T_\mathrm{ex}$, and $\int \tau dv$ the integrated optical depth of the absorption profile \citep[e.g.][]{mangum15}. For the OH 119\,\um doublet transitions, $E_l = 0$\,K, $A_{ul} = 0.138$\,s$^{-1}$ and $g_u = 6$ \citep{muller01,muller05}. Tabulated values of $Q_\mathrm{rot}$ are available from the NASA JPL spectroscopic database \citep{pickett98}. We assume an excitation temperature $T_\mathrm{ex} = 100$\,K as found in literature OH studies \citepalias[e.g.][]{gonzalezalfonso17}; the combination of $T_\mathrm{ex}$-dependent terms in Eq.~\ref{eq:NOH} varies by about a factor of 3 for 50$< T_\mathrm{ex} <$150. It is then straightforward to calculate the total (H) column density assuming an OH abundance, $N_\mathrm{H} = N_\mathrm{OH}/[\mathrm{OH}/\mathrm{H}]$. We adopt an OH abundance $[\mathrm{OH}/\mathrm{H}]=2.5 \times 10^{-6}$, as commonly assumed in the literature based on OH studies of the Milky Way star-forming region Sgr~B2 \citep{goicoechea02}.

In order to isolate only outflowing material, it is common to integrate the optical depth over a limited range of velocities. Here we calculate the integrated optical depth over velocities more blueshifted than $-$200\,\kms, a commonly-adopted threshold. Although this is not fast enough to be guaranteed to trace only outflowing material, we expect it to largely trace the outflows even in our sample DSFGs that often show broad CO or \cii emission line profiles \citepalias{spilker20}. Under the assumption of optically-thin absorption, the integrated optical depth over this velocity range is equal to the equivalent width over the same range. In practice, because the outflow rate itself is proportional to \vout (Eq. \ref{eq:mdot}), we instead calculate $\int_{-\infty}^{-200} \tau(v) \, v \, dv$ from our spectral fitting procedure, consequently incorporating the \vout term of Eq. \ref{eq:mdot} into the column density calculation directly. This allows us to include a first-order consideration of the shape of the absorption profiles while also removing the need to adopt a characteristic velocity in the outflow rate calculation. Finally, because we expect that the wind material does not fully cover the source, we adopt a covering fraction $\fcov = 0.3$, the average value determined for the low-redshift reference sample \citepalias{gonzalezalfonso17}. We discuss covering fractions in detail in \citetalias{spilker20}, where we estimate covering fractions ranging from a hard lower bound of $\sim$0.1 to upper limits of $\sim$0.7. These covering fractions are also not directly comparable: \citetalias{gonzalezalfonso17} estimate \fcov using their multi-transition OH radiative transfer analysis, while our estimates are based on our lensing reconstructions with hard lower limits based on spectral analysis. While assuming a different value for \fcov would linearly rescale our optically-thin outflow rate estimates, this has no impact on our subsequent results, as we explain further below.

\begin{figure*}
\begin{centering}
\includegraphics[width=0.8\textwidth]{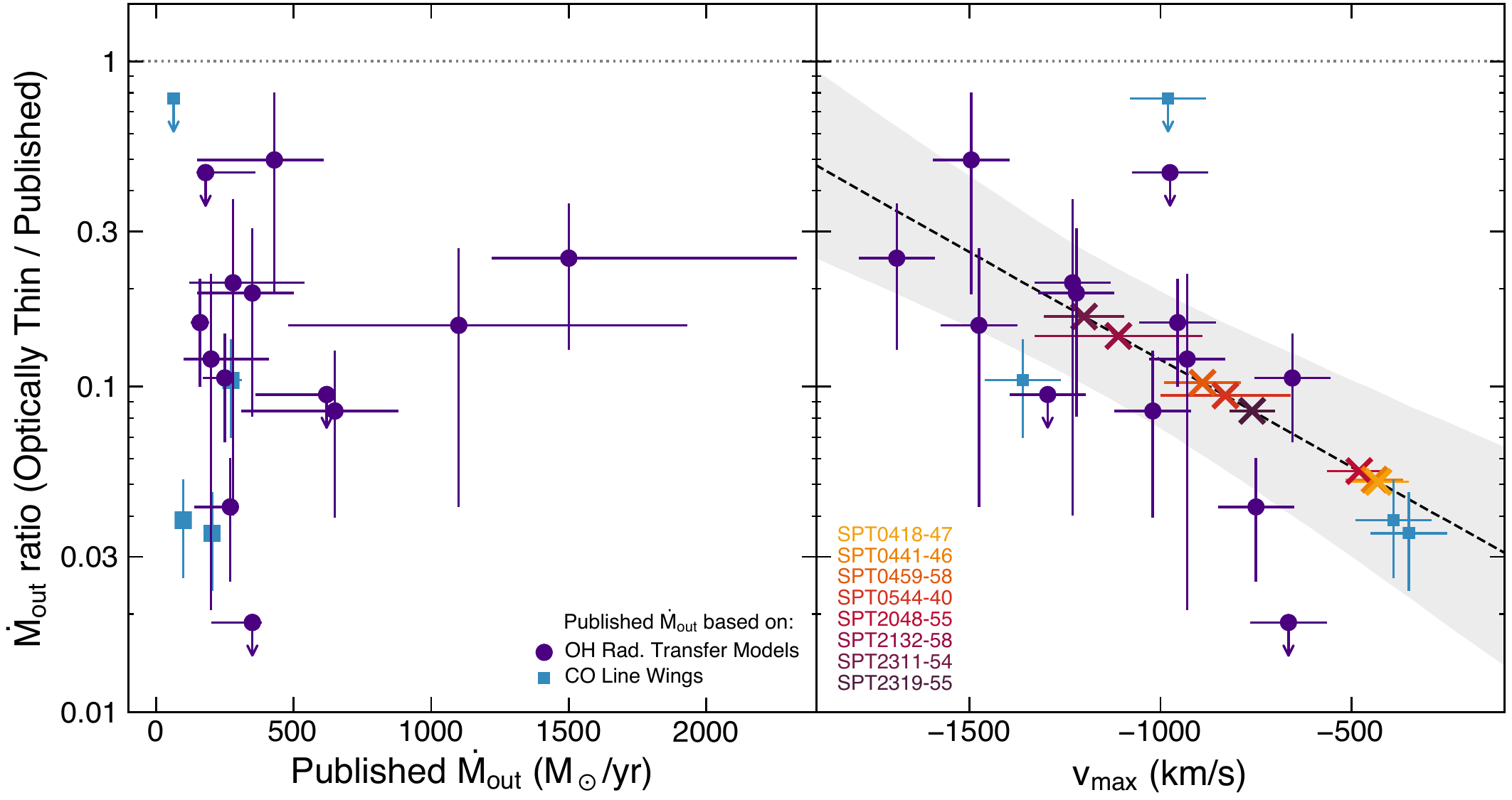}\\
\end{centering}
\caption{
The ratio of optically thin outflow rates to the published values for the literature reference sample (blue symbols) against the published outflow rates (left) and the maximum outflow velocity \vmax (right). Literature objects with outflow rates derived from multi-transition OH radiative transfer are shown with circles, while those with only CO-based rates are shown as squares. Outflow rates assuming the OH 119\,\um absorption is optically thin underestimate the true outflow rates by an amount that is correlated with the outflow velocity. The right panel shows a log-linear fit and 16-84th percentile confidence interval (including an intrinsic scatter of $\pm$0.15dex) that we use to `correct' the optically thin outflow rates to more realistic estimates using the measured values of \vmax for the high-redshift SPT sources ($\times$ symbols). The SPT sources in the right panel are placed along the best-fit line according to their measured \vmax.
}\label{fig:optthin}
\end{figure*}

We use Eq.~\ref{eq:mdot} to calculate the outflow rates for our own and the literature reference samples. For the reference sample, we use far-IR continuum sizes from PACS 100\,\um imaging \citep{lutz16}, or assume the average size $\approx$1\,kpc if no data are available. We derive minimum optically thin outflow rates spanning $8-370$\,\Msol/yr for the literature sources and $5-120$\,\Msol/yr for the SPT sample, and corresponding outflow masses $\log(\Mout/\Msol) \approx 7 - 8.5$. We emphasize that these values are strong lower limits given the expected high OH line opacities.

Figure~\ref{fig:optthin} (left) shows the ratio of the optically thin outflow rates to the published values from the OH radiative transfer models and CO line wings for the literature sample; upper limits in this plot correspond to those sources that were spatially unresolved by PACS and therefore have upper limits on \Rout. As expected, the optically thin assumption likely underestimates the true outflow rate by a large factor, $\approx 4-30 \times$ for most sources.  The fact that the outflow rates can be so drastically underestimated is at some level a testament to the sensitivity of OH 119\,\um to even minute amounts of outflowing material: with ALMA at $z>4$, in principle it is possible to detect molecular outflow rates of just $\sim$10\Msol/yr in less than an hour of observing time. This is a consequence of both the relatively high OH abundance and especially the large value of the Einstein $A_{ul}$ for the ground-state 119\,\um transition, $\sim$10$^5$ times larger than for \cii 158\,\um or $\sim$10$^6$ times larger than for low-order CO transitions. The drawback to this sensitivity, of course, is that the line opacities are high and it is not easy to determine by what exact factor the true outflow rate has been underestimated, as Figure~\ref{fig:optthin} shows.

There is no obvious trend between the ratio of optically thin to published outflow rate and the published outflow rate itself; evidently the OH 119\,\um line opacity varies by a large amount in different galactic winds. We do, however, identify correlations between this ratio of outflow rate estimates and various measures of the outflow velocity. Sources with the fastest outflows are also the closest to being consistent with optically thin absorption. Figure~\ref{fig:optthin} (right) shows this in terms of \vmax, but we obtain results consistent within the uncertainties from \vef and \vfifty as well; while \vmax is somewhat more difficult to measure \citepalias{spilker20}, it shows the largest dynamic range among the outflow velocity metrics. Such a correlation makes intuitive sense: for a given column density of absorbing gas, if the total gas column extends over a larger range in velocity, the line opacity per unit velocity interval must necessarily be lower and therefore the line opacity averaged over the full absorption profile must also be lower. This leads to a less extreme `correction factor' needed for sources with very fast outflows.

We fit a simple log-linear function to the data in Figure~\ref{fig:optthin} (right), determining uncertainties on the fit using the same Monte Carlo resampling method we use for all outflow rate techniques. We find a best-fit relation for the outflow rates `corrected' from the optically-thin values of
\begin{equation}
\log(\Mdot^{\mathrm{thin\,corr.}}/\Mdot^{\mathrm{thin}}) = m(\vmax + 1000) + b, 
\end{equation}
with $m = -6.4^{+1.8}_{-1.7} \times 10^{-4}$\,(\kms)$^{-1}$ and $b = 0.91^{+0.07}_{-0.06}$, and \vmax in \kms. The analysis indicates an intrinsic dispersion of $\sim$0.15\,dex around the best-fit relation in addition to the statistical uncertainties. We use this relation and the measured values of \vmax to estimate the true outflow rates empirically corrected from the optically-thin assumption. 

We note again that our prior assumption of $\fcov = 0.3$ in our calculation of $\Mdot^{\mathrm{thin}}$ has no impact on our `corrected' outflow rates, as a different assumed value propagates directly into $b$ in the equation above. There is also no evidence for a correlation between \fcov and other galaxy properties in the low-redshift training sample that could influence our outflow rates given the differences between the samples \citepalias{gonzalezalfonso17}. While we do find a tentative correlation of \fcov with \lir \citepalias{spilker20}, we expect those covering fractions to be upper limits on the true values and stress again that the methods used between low- and high-redshift are not directly comparable. Our assumed $\fcov = 0.3$ lies well within the lower and upper limits we expect for the true values, so we do not expect this to add substantial additional uncertainty beyond the present estimates. Both the optically thin and the corrected outflow rates are given in Table~\ref{tab:sptmdots}.

\subsection{Simple Empirical Estimates} \label{simpempirical}

As we expected, the optically thin outflow rates almost certainly severely underestimate the true outflow rates. While we derived a method to correct these values to more realistic outflow rates, the correction factors remain highly uncertain and in any case the general methodology deserves to be cross-checked by other methods. We now consider two simple empirical methods to provide alternative estimates of the outflow rates before moving to a more complex empirical method.

In \citetalias{spilker18a} we made a simple estimate of the true outflow rates using a subset of the present literature reference sources for which outflows had also been detected in CO emission. In that work we took the OH 119\,\um equivalent widths for the low-$z$ literature sources integrated over the velocity ranges where high-velocity wings of CO emission had been detected \citepalias{gonzalezalfonso17}, under the philosophy that both traced molecular outflows and that the outflows should appear over the same velocity range in both tracers. Here we follow a similar vein, now including an expanded reference sample. Instead of individually choosing velocity ranges over which to measure the OH equivalent widths, here we simply fit a linear relationship between the published literature \Mdot values and \EWvoh from our re-measured 119\,\um spectral fits, which gave equivalent widths similar to those we used in \citetalias{spilker18a}. 

\begin{figure*}
\begin{centering}
\includegraphics[width=0.8\textwidth]{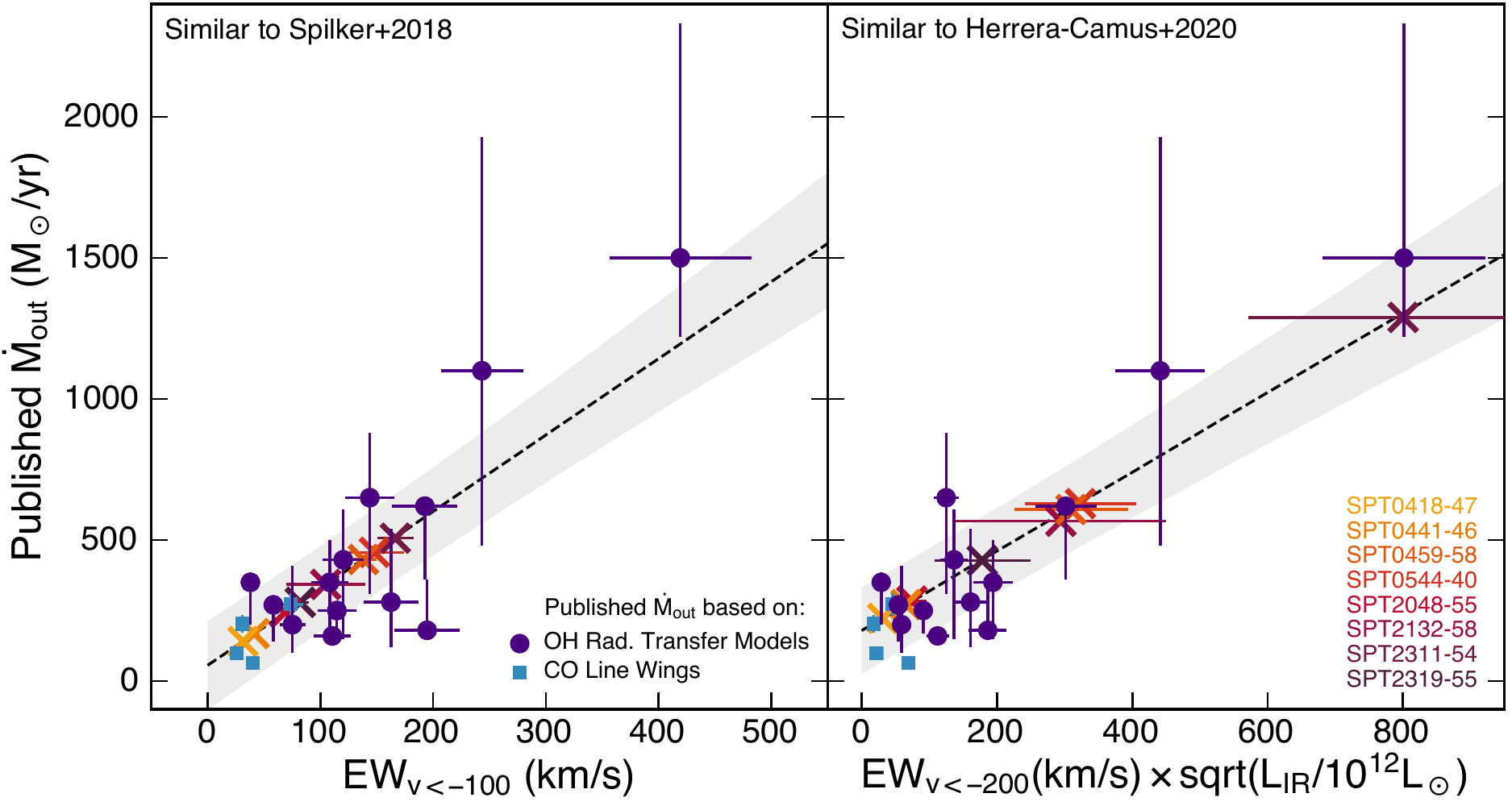}\\
\end{centering}
\caption{
Two simple empirical ways we use to estimate molecular outflow rates, using parameterizations in the style of \citet{spilker18a} (left, assuming \Mdot is correlated with the blueshifted equivalent width alone) and \citet{herreracamus20} (right, assuming an additional dependence on $\lir^{1/2}$). 
Each panel also shows a linear fit and 68 percent confidence interval (including an intrinsic scatter of $\pm$150\,\Msol/yr; dashed line and grey shaded region). We use these fits and the measured OH 119\,\um spectral properties to infer outflow rates for the high-redshift SPT sources. All symbols as in Fig.~\ref{fig:optthin}.
}\label{fig:empS18HC20}
\end{figure*}

Figure~\ref{fig:empS18HC20} (left) shows the results of this analysis. We do not force this fit to have a zero intercept. Although this allows the unphysical scenario of positive outflow rates in the absence of any absorption or even negative outflow rates for low equivalent widths, allowing this freedom in the model yields a better characterization of the uncertainty at low \Mdot. We find a best-fit expression for the outflow rate 
\begin{equation}
\Mdot^{\mathrm{S18}} = m \EWvoh + b,
\end{equation}
with $m = 2.7^{+0.4}_{-0.5}$\,\Msol/yr/(\kms) and $b = 55\pm50$\,\Msol/yr, the outflow rate in \Msol/yr and the equivalent width in \kms. We find an intrinsic dispersion of $\pm$150\Msol/yr around this relation in addition to the statistical uncertainties that at least applies in the low-\EWvoh, low-\Mdot regime. At higher \Mdot there are too few sources to quantify any additional scatter beyond the statistical uncertainties; we assume a constant 150\,\Msol/yr scatter for all values of \EWvoh.

The OH equivalent width is not expected to be the sole controlling parameter that predicts outflow rates, of course. \citet{herreracamus20} (abbreviated \citetalias{herreracamus20}) explored an alternative simple parameterization, fitting the outflow rates to the product $\EWvth \times \sqrt{\lfir}$. This was motivated by an expectation that the outflow rate should depend on both the column density of outflowing gas (related to \EWvth) and the size of the source (proportional to $\sqrt{\lfir}$ through a Stefan-Boltzmann type relation), as in Eq.~\ref{eq:mdot}. 

We repeat a similar analysis as \citetalias{herreracamus20}, with a couple small modifications. First, we use $\sqrt{\lir}$ instead of $\sqrt{\lfir}$, which is more readily available for all literature reference sources. Second, we do not fit a line forcing the y-intercept to be zero as done in \citetalias{herreracamus20}. Again, this allows us to better understand the uncertainties at low \Mdot. Our best-fit relation for the outflow rate in this way is
\begin{equation}
\Mdot^{\mathrm{HC20}} = m (\EWvth \sqrt{\lir/10^{12}\Lsol}) + b,
\end{equation}
again with the outflow rate in \Msol/yr and \EWvth in \kms, $m = 1.40^{+0.21}_{-0.25}$\,\Msol/yr/(\kms) and $b = 180^{+40}_{-30}$\,\Msol/yr. We find an essentially identical intrinsic scatter around this relation as before, $\approx$150\,\Msol/yr, where this is assumed to be constant due to the lack of sources with very high outflow rates. 

Aside from a more physically-justified parameterization, this method also has a slightly higher dynamic range in the abscissa than the \citetalias{spilker18a}-style fit. Between the two methods, we have some preference for the \citetalias{herreracamus20} parameterization. Outflow rates derived from both methods are given in Table~\ref{tab:sptmdots}.

\subsection{Multivariate Empirical Estimate} \label{plsempirical}

Finally, we consider a more complex empirical model to derive outflow rates. While the analyses in the previous subsection relied on specific linear correlations between observables and published outflow rates, there is no particular reason to choose those specific observables over others apart from some physical intuition about the likely important parameters. The fact that we find significant additional intrinsic scatter beyond the inferred uncertainties in the previous fits is a clue that a more complex model connecting the observables and the outflow rates is warranted. Indeed, both our reference and high-redshift samples have many more known properties than we have yet utilized, both from the OH spectra themselves as well as ancillary measurements from other data. Here we perform one final analysis that attempts to discern the most predictive relationship between all available measurements and the outflow rates, at the expense of linking the resulting relationship to any particular physical meaning.

To explore the complex relationship between outflow rates and all available measurements, we use a `partial least squares' (PLS) technique \citep{wold66}. PLS is both a regression and dimensionality reduction technique, and can be thought of as somewhat of a hybrid between standard multivariate linear regression and principal component analysis (PCA) or singular value decomposition. PLS is well-suited to cases such as ours where the number of objects in the reference sample is relatively few but the number of measured quantities for each sample object is large, with many of the measured quantities correlated with each other. In our case, for example, while \vef and \vmax encode slightly different information about the shape of the OH absorption profile, they are still strongly correlated: fast outflows are fast regardless of the metric used. While PCA techniques are capable of describing the variance in the observables, not all principal components need be predictive of some other quantity (\Mdot in our case). PLS addresses this by maximizing the covariance between the space of observables and the space of desired predicted quantities. PLS also performs better than some other techniques (e.g. random forest estimators) when some measured observables of the target sample (in this case the SPT objects) lie outside the dynamic range of the training sample -- that is, when extrapolation is required for one or more observables. In the end this has little influence on our application because the most predictive observables (see below) are well-sampled by the reference objects and extrapolation is not generally required.

\begin{figure}
\begin{centering}
\includegraphics[width=\columnwidth]{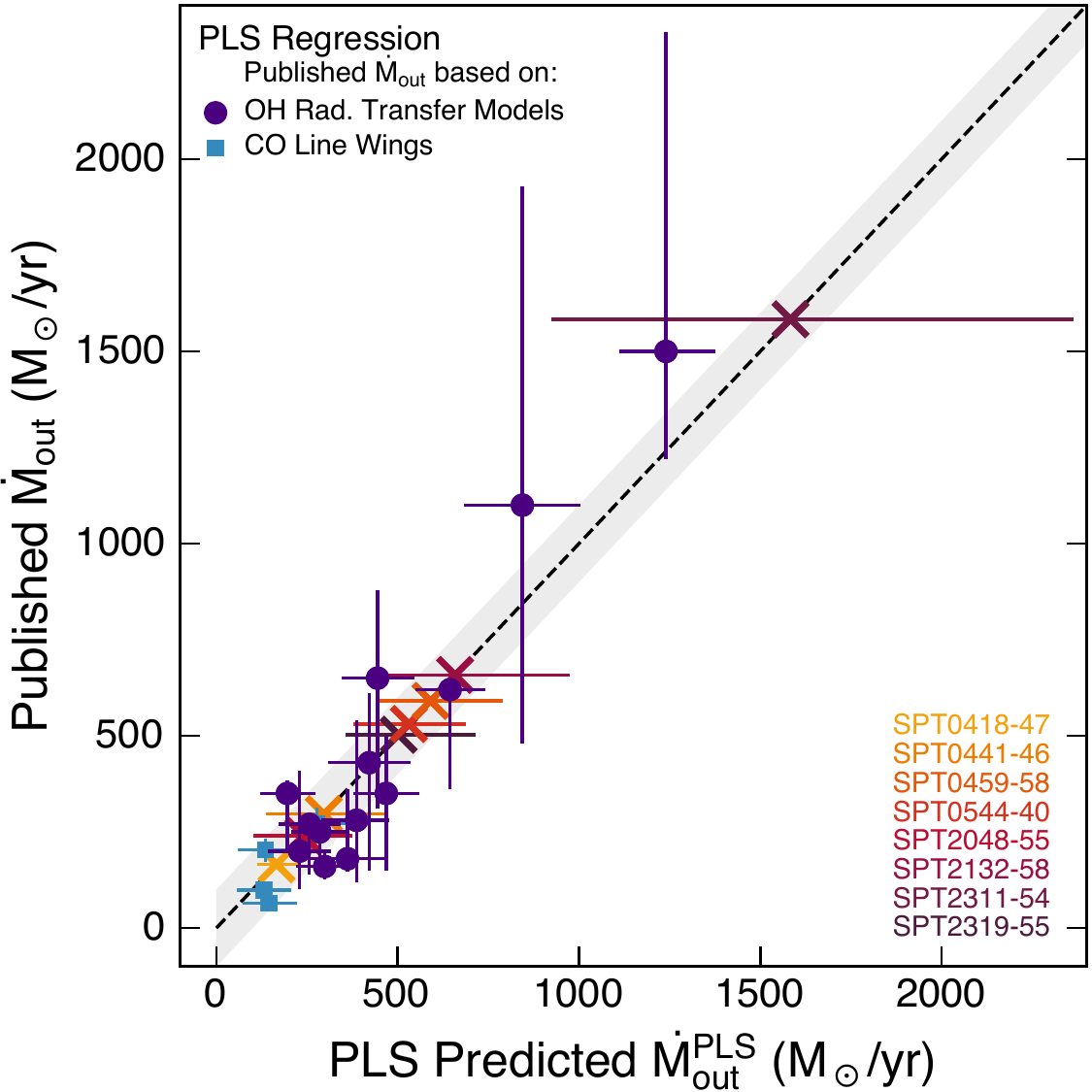}\\
\end{centering}
\caption{
Predicted and measured outflow rates from the empirically-based PLS method that does not presuppose any particular functional form between measured properties and the molecular outflow rate. Partial least squares (PLS) is a technique that combines dimensionality reduction with multivariate regression; see Section~\ref{plsempirical}. The dashed line shows the one-to-one relation while the grey shaded region shows the upper limit to the remaining intrinsic scatter, $\pm$100\,\Msol/yr.
}\label{fig:pls}
\end{figure}

For our purposes, we use PLS to predict the outflow rates \Mdot from a variety of (sometimes strongly correlated) observed properties: several metrics of the OH velocity profiles and equivalent widths integrated over various velocity ranges, as well as ancillary galaxy properties such as \lir, \rdust, the AGN contribution to the bolometric luminosity \fagn, and the effective dust temperature \Tdust. We experimented extensively with various numbers and combinations of observables and found consistent results for the predicted outflow rates of the SPT sources in almost all cases. Generally regardless of the observables used, a maximum of four or five PLS components minimized the mean squared error in the predicted outflow rates of the reference sample (that is, the dimensionality of the problem could be reduced from the number of observables used to four or five, due to covariances between the observables employed). PLS also allows us to understand which observables are most responsible for driving predictions for the outflow rates. Of those we explored, the outflow velocity and equivalent width were the most predictive of the measured outflow rates, while \fagn and $r_\mathrm{dust}$ generally had little predictive power, possibly due to the relatively small dynamic range in these quantities in the training and target samples. 

Figure~\ref{fig:pls} shows the comparison between predicted and published outflow rates for the combination of parameters that includes \vfifty, \vef, \vmax, \EWvoh, \EWvth, \EWtot, \lir, and \fagn. Interestingly, unlike the previous methods, there is no longer any detectable intrinsic scatter between the predicted and published outflow rates; the grey shaded region in Figure~\ref{fig:pls} illustrates the approximate upper limit on the scatter we can set with the data available, $\approx$100\,\Msol/yr. Although it remains rather unsatisfying to necessarily discard all physical interpretation of the resulting predictions, clearly PLS is capable of translating the complex measurement space into the desired output outflow rates. Predicted outflow rates from this method are provided in Table~\ref{tab:sptmdots}.

\subsection{Summary and Method Comparison} \label{mdotcompare}

\begin{figure*}
\begin{centering}
\includegraphics[width=\textwidth]{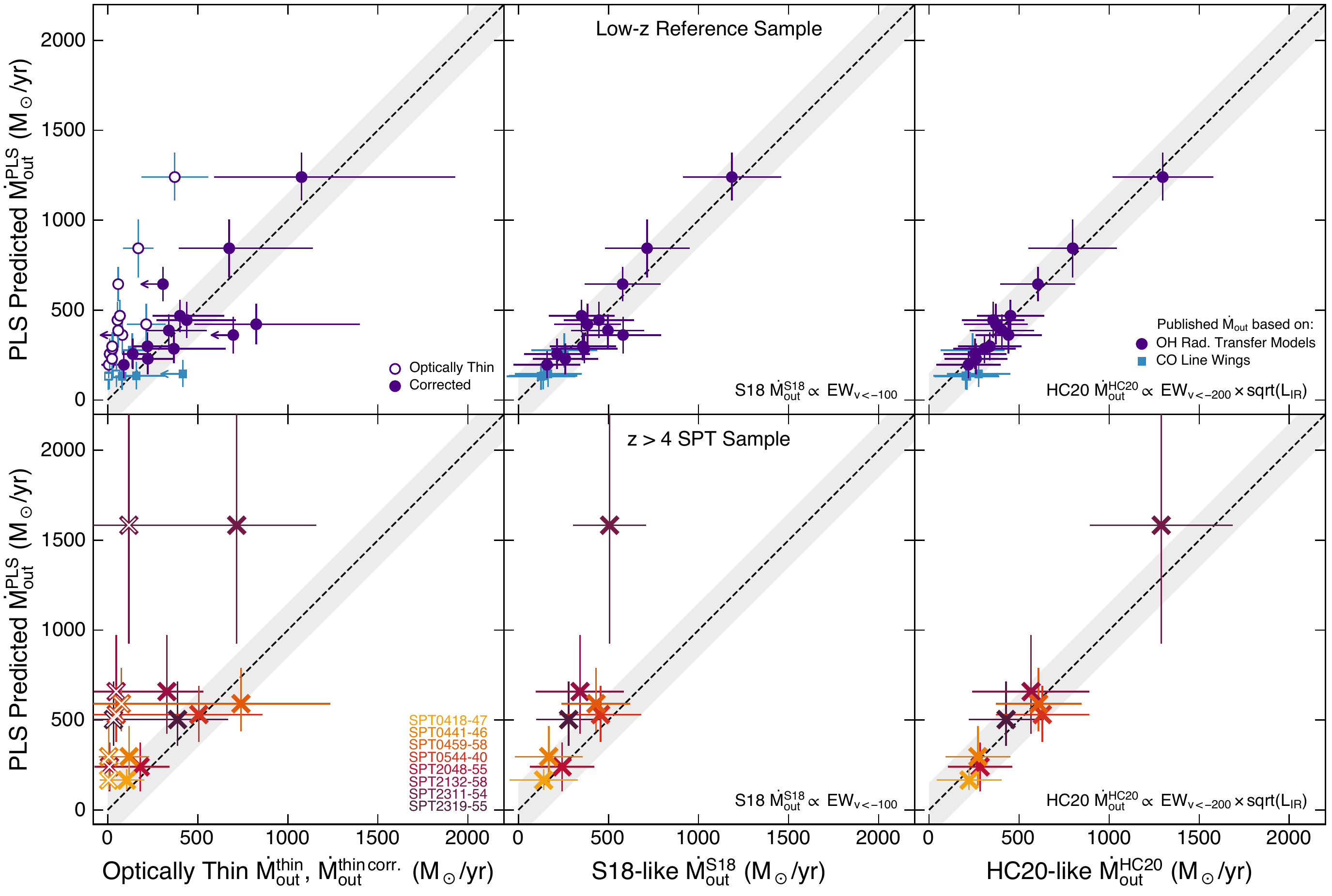}\\
\end{centering}
\caption{
Comparison of all outflow rate estimations we have used. For easier visualization, the literature reference sources are along the top row and the inferred values for the SPT sample on the bottom row. To guide the eye, dashed lines in each panel show the one-to-one line while the grey shaded region shows $\pm$150\,\Msol/yr about this relation. The y-axis shows the rates derived from the multivariate empirical PLS technique (Sec.~\ref{plsempirical}), while the columns show the results from an optically-thin estimate (left; Sec.~\ref{optthin}) and two simple empirical methods similar to those from \citetalias{spilker18a} (center) and \citetalias{herreracamus20} (right; Sec.~\ref{simpempirical}). In the left-hand column, open symbols are the optically-thin outflow rates, while the filled symbols apply the `correction' from Section~\ref{optthin}. 
}\label{fig:mdotcompare}
\end{figure*}

We now have four different estimates for the molecular outflow rates applied to the high-redshift SPT objects -- one corrected from the optically-thin assumption, two simple empirical estimators, and one more complex empirical estimate. Figure~\ref{fig:mdotcompare} compares these estimates for both the low-redshift reference sample and as applied to our $z>4$ objects. Note that for the literature sources this Figure only compares the outflow rates predicted from each method to each other; comparisons with the `true' published values can be found in the preceding Figures.

Essentially by definition this Figure shows good agreement between the various methods for the reference sample, since this sample was used to derive the conversions between observables and outflow rates in the first place. We also find generally good agreement between the estimators for the SPT objects, in particular between the multivariate PLS analysis and the simpler approach of \citetalias{herreracamus20}. Evidently these methods make use of the most salient predictive measurements from the OH spectra and ancillary galaxy properties.

\begin{figure*}
\begin{centering}
\includegraphics[width=\textwidth]{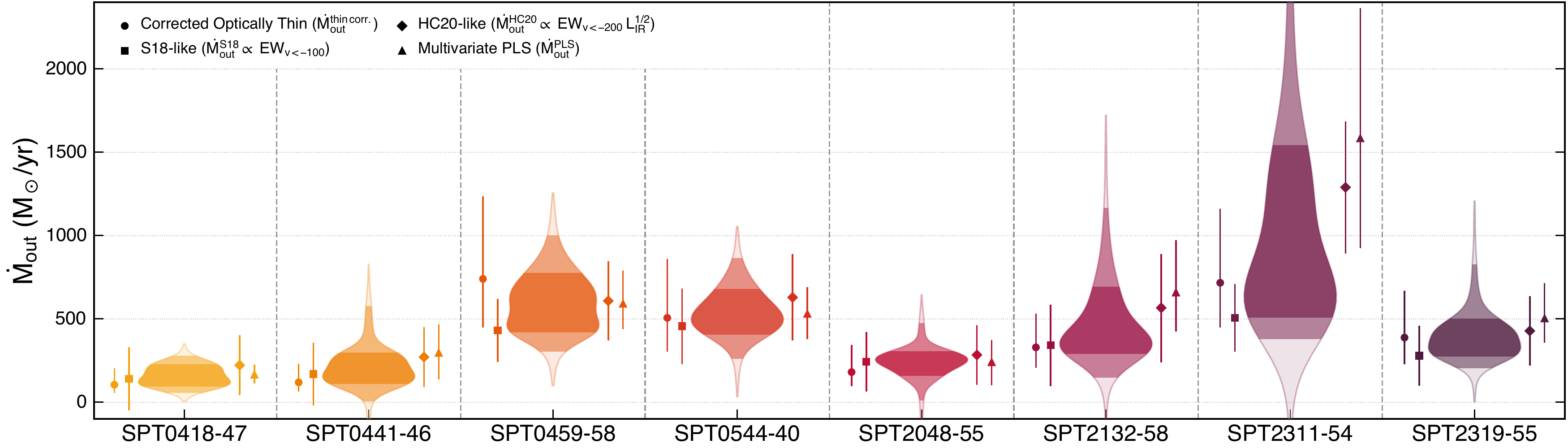}\\
\end{centering}
\caption{
Summary of outflow rates for the high-redshift SPT sources. For each object in the sample, the symbols show the outflow rates derived from each method as indicated, while the violin plot shows the joint distribution from all four methods (weighted equally). Within the violin plots, the darker shaded regions indicate the 68 and 95 percent confidence intervals.
}\label{fig:sptmdotall}
\end{figure*}

Figure~\ref{fig:sptmdotall} compares the outflow rates for the SPT sources in more detail. This figure shows the outflow rates derived from each method for each source. This figure again demonstrates the generally good agreement between methods, although the PLS and \citetalias{herreracamus20}-like methods tend to yield slightly higher values than the other methods. We also show a joint distribution of the outflow rates created by equally combining the Monte Carlo trials from each method. While this should not be considered a true joint probability distribution of the outflow rates -- the methods to derive input distributions are hardly independent, for example -- it both highlights the level of agreement or disagreement between methods and summarizes the constraints we place on the outflow rates. These joint outflow rates are listed in Table~\ref{tab:sptmdots}, referred to as $\Mdot^{\mathrm{joint}}$. We use these joint estimates and associated uncertainties throughout the remainder of the text as our `best' estimates of the outflow rates, subsequently dropping the `joint' superscript from the notation for simplicity.

The joint distributions from each method suggest that we are able to estimate the outflow rates for our sources at about the factor-of-two level. The OH-based reference sources have quoted uncertainties at the $\sim$50\% level; the higher level of uncertainty for our sources reflects the lack of additional OH data for our sample that propagates into the scatter seen in the four individual methods and thus into our final joint estimates. 

We emphasize that for both samples these uncertainties are likely underestimated due to systematics in many of the assumptions, from the OH abundance to the assumed geometry and outflow history. For our high-redshift objects, while our estimates are empirically based, the methods we have described presume that low-redshift IR-luminous galaxies are sufficiently similar to our targets as to not render these calculations meaningless. While the observed characteristics of our sample are contained within the parameter space probed by the reference sample (Figures~\ref{fig:optthin}--\ref{fig:pls} and \citetalias{spilker20}), it is certainly possible that some other unmeasured quantity has a strong influence on the outflow rates that is not accounted for by our methods. Thus while we propagate the uncertainties on the joint outflow rates in the remainder of the text, it is important to remember that these are probably more uncertain by some difficult-to-quantify amount.

\section{Results and Discussion} \label{results}

\subsection{Outflow Driving Mechanisms} \label{driving}

In low-redshift samples, correlations between outflow velocities and host galaxy properties such as SFRs or AGN luminosities have been used to shed light on the physical mechanism(s) responsible for launching the outflows. There are good theoretical reasons to believe that the energy and momentum imparted to the gas from star formation and/or AGN activity should play a role in driving galactic winds, and should then manifest in the properties of the outflows launched. Nevertheless, observations of neutral and low-ionizations species show at best weak correlations between outflow velocities and SFR from the local universe to $z\sim1$ \citep[e.g.][]{weiner09,rubin14,chisholm15,robertsborsani19}, with any trend mostly due to the weak outflows seen in very low-SFR galaxies \citep[e.g.][]{heckman16}. 

While this could plausibly be because the neutral outflows are less strongly coupled to the driving source, similarly weak correlations have also been seen for the molecular phase traced by OH in nearby ULIRGs and QSOs \citep[e.g.][]{veilleux13}. While still relatively weak, the strongest correlations between outflow velocities and galaxy properties are found with \LAGN and \fagn, suggesting a connection between AGN and wind launching at least in these extreme nearby systems.

\begin{figure*}
\centering
\includegraphics[width=\textwidth]{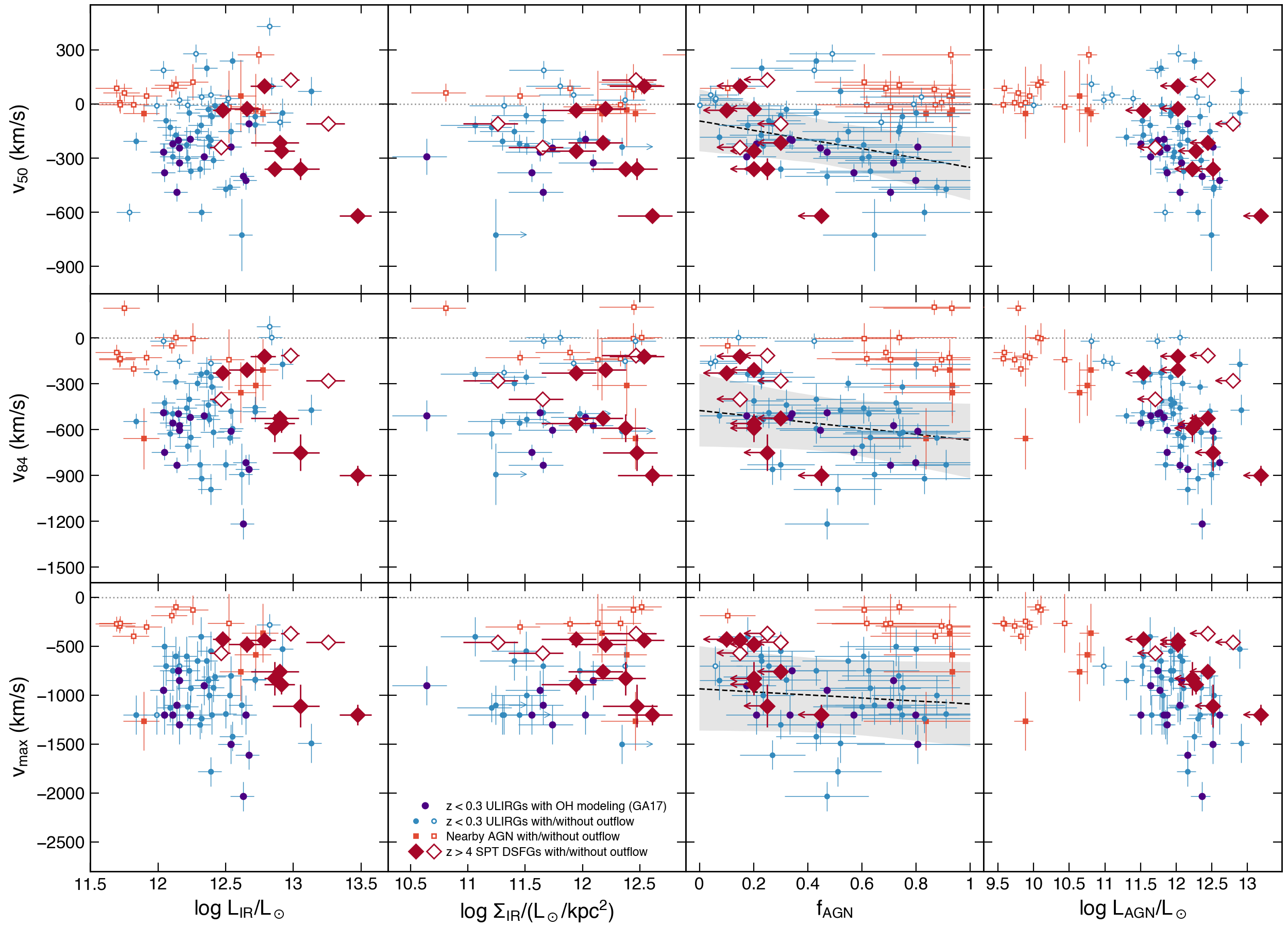}
\caption{
Outflow velocity metrics as a function of \lir, \sigir, \fagn, and \LAGN. All velocities are measured from OH 119\,\um absorption spectra. Red diamonds show our high-redshift sample, blue circles show the combined sample of nearby ULIRGs and QSOs (\citealt{spoon13,veilleux13,calderon16,herreracamus20}; see \citetalias{spilker20} for details), and orange squares nearby AGN-dominated galaxies \citep{stone16}. Filled symbols indicate sources with outflows and empty those without, as determined by the original authors of each study. The nearby ULIRGs with OH-based radiative transfer models to measure outflow rates (Section~\ref{litref}) are highlighted as larger navy circles. Previous low-redshift work indicated that \fagn is correlated with the outflow velocities, so in the third column we show simple linear fits with 68 percent confidence intervals to the low-redshift objects with outflows. While we currently have no evidence of AGN activity in the high-redshift SPT sample, our objects are not obvious outliers in these plots, suggesting that we cannot rule out AGN as the driving mechanism of the outflows we have observed.
}\label{fig:kinematics}
\end{figure*}

Figure~\ref{fig:kinematics} shows three outflow velocity metrics, \vfifty, \vef, and \vmax, as a function of \lir, the IR surface density \sigir, \fagn, and \LAGN, where we now compare our high-redshift objects to the combined sample of low-redshift ULIRGs and QSOs and nearby AGN-dominated systems (described in detail in \citetalias{spilker20}). We distinguish between objects with outflows (filled symbols) and those without (empty), as determined by the original authors. We also note that the subset of low-redshift ULIRGs selected for OH radiative transfer modeling by \citetalias{gonzalezalfonso17} is skewed towards sources with the fastest outflows, presumably because these were a more viable sample for multi-transition modeling. We return to this point several more times because it propagates into many of the differences we see with the low-$z$ ULIRGs in other outflow properties as well.

The left column of Fig.~\ref{fig:kinematics} first shows outflow velocities as a function of \lir. In agreement with \citet{veilleux13}, we see no evidence of a correlation in the expanded sample of nearby galaxies. Interestingly, however, we do see hints of a trend within the $z>4$ SPT DSFGs when considered alone, with the most luminous sources also driving the fastest outflows. Whether this is a genuine difference between the outflows driven in low- and high-redshift galaxies remains to be seen; a larger sample of high-redshift objects that spans a wider range in \lir and other properties will be required to understand these tentative differences further.

Instead of \lir alone one might instead expect the outflow velocity to depend more strongly on the IR surface density \sigir (or similarly the SFR surface density), for example in cases in which radiation pressure on dust grains drives the outflows \citep[e.g.][]{thompson15}. The second column of Fig.~\ref{fig:kinematics} shows these quantities for the low- and high-redshift OH molecular outflow samples, where we have used far-IR sizes measured from \textit{Herschel}/PACS imaging \citep{lutz16,lutz18} for the low-redshift samples and the sizes from our lensing reconstructions for the SPT sample \citepalias{spilker20}. We find no convincing evidence of correlation between these quantities even for the SPT sample considered alone.

Of the parameters investigated by \citet{veilleux13} for low-redshift ULIRGs and QSOs, the strongest correlations with outflow velocities were found with \fagn and \LAGN,\footnote{These correlations excluded the most AGN-dominated systems where OH was seen purely in emission; none of our sample shows OH in emission either purely or partially.} which those authors argued could be due to obscuration effects whereby the fastest-moving material was more easily visible in AGN that had already cleared the nuclear regions or were oriented face-on. The subsequent addition of far less luminous AGN-dominated systems by \citet{stone16} agreed with this picture although the number of sources with definite outflows was small. The third column of Fig.~\ref{fig:kinematics} shows outflow velocities against \fagn. We also fit a simple linear function to the low-redshift sources with molecular outflows, finding a marginally significant correlation with \vfifty that becomes weaker with \vef and \vmax; the scatter is clearly large. The limits on \fagn for the SPT sources based on rest-frame mid-IR photometry do not clearly result in these objects being outliers, and they certainly would not be outliers even if we have underestimated \fagn by a substantial amount (Section~\ref{outflowratemethods}).

Finally, the right column of Fig.~\ref{fig:kinematics} shows outflow velocities as a function of \LAGN, which \citet{stone16} find to be strongly correlated in low-redshift sources in agreement with \citet{veilleux13}. We also see some relationship between these quantities -- namely, sources with low AGN luminosities rarely drive fast outflows. However, we note that while the \citet{stone16} sample certainly extends the dynamic range in \LAGN probed, this now conflates samples selected in very different ways, with many other possible confounding variables (mass, for example). Regardless, we again find that the limits we can place on \LAGN for the SPT sample again do not make them obvious outliers.

In summary, among the SPT DSFGs alone, the total \lir appears to be most strongly correlated with outflow velocity, although a larger sample size will be required to investigate whether this is genuine.  While we recover correlations previously noted in low-redshift work with our larger combined literature sample, the OH outflow velocities appear to be at best weak indicators of the driving source of molecular outflows, with substantial scatter. While we currently have no evidence for AGN activity in the SPT DSFGs and only weak limits on \fagn, our objects are not obvious outliers in plots of outflow velocity and AGN properties given the substantial scatter seen amongst the low-redshift objects, and we thus cannot rule out that AGN are responsible for driving the molecular outflows we have observed.

\subsection{Molecular Outflow Rate Scaling Relations} \label{massloading}

A number of recent works have explored scaling relations between molecular outflow properties and host galaxy properties, compiling samples of now dozens of objects \citep[e.g.][]{cicone14,gonzalezalfonso17,fluetsch19,lutz20}. While these studies focused exclusively on low-redshift galaxies, we now include our measurements for the first sample of molecular outflows in the early universe. Our primary comparison samples are the OH-based outflow measurements in nearby ULIRGs from \citetalias{gonzalezalfonso17}, as before, supplemented with the CO-based sample of \citet{lutz20}, which extends to lower-luminosity systems. All samples assume the same outflow geometry of Section~\ref{outflowgeom}. We note that \citet{lutz20} found that OH-based outflow rates tended to be $\approx$0.5\,dex higher than CO-based rates in their comparison of galaxies observed in both tracers (while the total outflow masses \Mout were very similar). Because the CO-based sample spans a different range of parameter space than the other samples, we also comment on how our inferences in this section would change if the CO outflow rates were increased by 0.5\,dex. We also detail changes to our interpretation that would result from doubling our present upper limits on \fagn to try to account for the effects of any heavily-obscured AGN that may not be detectable even in the rest-frame mid-IR. It is important to note that none of these samples at any redshift are complete or unbiased; the galaxies typically targeted for molecular outflow observations are highly biased towards luminous star-forming systems and/or quasars.

\begin{figure}
\begin{centering}
\includegraphics[width=\columnwidth]{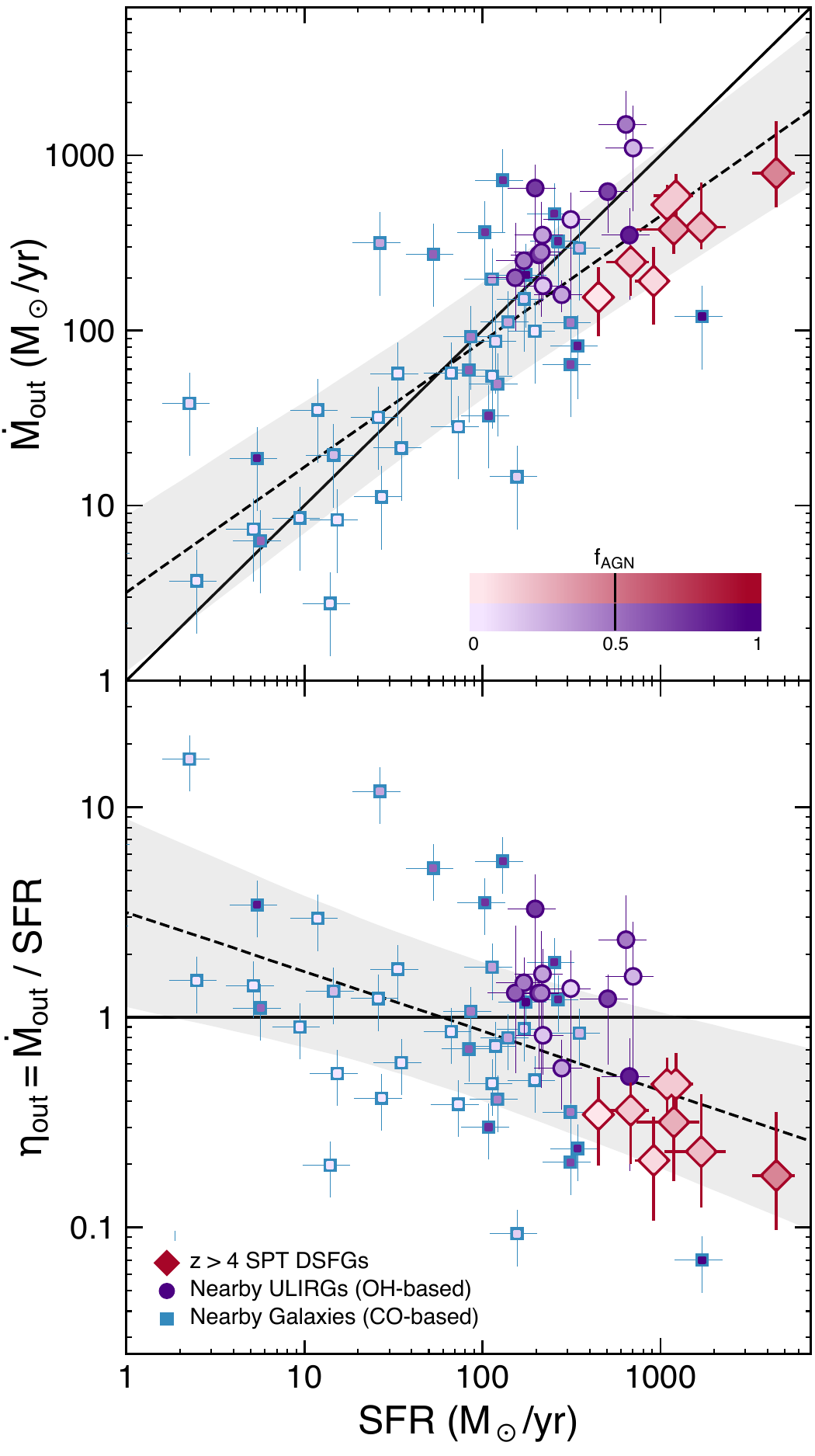}\\
\end{centering}
\caption{
Molecular outflow rates (upper panel) and mass loading factors (bottom) as a function of SFR for the high-redshift SPT DSFGs (diamonds), nearby ULIRGs with OH-based outflow rates (circles), and an assortment of nearby galaxies with CO-based outflow rates (squares). Points are colored by \fagn (or the limit on \fagn, for the SPT sources). Solid lines indicate the one-to-one relation in the upper panel and $\etaout=1$ in the lower panel. In both, dashed lines and grey shaded regions show the median and 68 percent confidence interval on power-law fits to the combined samples.
}\label{fig:mdotoutsfr}
\end{figure}

Figure~\ref{fig:mdotoutsfr} shows the molecular outflow rate \Mdot and mass loading factor $\etaout \equiv \Mdot/$SFR as a function of SFR. We find uniformly sub-unity mass loading factors for the high-redshift DSFGs, although the uncertainties of course remain significant. We would still find loading factors $\lesssim 1$ even if we have underestimated our limits on \fagn by a factor of 2 (which would consequently lower the SFR). This is a perhaps surprising result -- these galaxies are among the most luminous, highest-SFR objects known, yet drive relatively weaker outflows than many less-luminous nearby galaxies (though again, none of these samples is complete or unbiased). In particular, despite SFRs a few times higher than the low-redshift OH sample of \citetalias{gonzalezalfonso17}, the outflow rates we derive do not increase accordingly and the loading factors are consequently lower. At least some of this difference is likely due to the selection for fast outflows in the low-redshift work, but because the outflow rate depends only linearly on the velocity this is insufficient to explain the full difference.

Our sample does appear, however, to follow the slightly sub-linear relationship seen in some low-redshift studies \citep[e.g.][]{fluetsch19}, extended now to an order of magnitude higher SFR and the high-redshift universe. Fitting a power-law to the combined samples, we find a best-fit relationship $\log(\Mdot) = (0.72\pm0.05)\log(\rm{SFR}) + (0.5\pm0.1)$, with \Mdot and SFR in \Msol/yr and an additional intrinsic scatter on the outflow rates of $\approx$0.25\,dex. Thus, we find a transition from $\etaout > 1$ to sub-unity values near SFR$\sim$100\,\Msol/yr. Because of the distribution of the CO-based \citet{lutz20} sample in SFR, increasing the CO-based outflow rates by 0.5\,dex would further flatten the power-law slope to $\approx$0.5 but increase the transition SFR at $\etaout=1$ to 500\,\Msol/yr. On the other hand, if we lowered the SFRs of our sample by doubling \fagn to estimate the effect of possible highy-obscured AGN, the power-law slope would marginally increase to $\approx0.8$.

Interestingly, we find no evidence that the dominance of an AGN plays a secondary role in setting the outflow rate, once the overall trend with SFR (or \lir) has been accounted for. The points in Fig.~\ref{fig:mdotoutsfr} are color-coded by \fagn, and it is clear that the remaining scatter in the \Mdot--SFR relationship is uncorrelated with \fagn. This remains a point of some contention in the recent literature: \citet{cicone14} and \citet{fluetsch19} do claim a correlation between \etaout and \fagn for $\fagn\gtrsim0.7$, while \citet{lutz20} find no such correlation, even though both studies use a largely-overlapping set of literature outflow detections (since we use the \citealt{lutz20} CO-based literature compilation it is no surprise that we also find no correlation given the fairly small increase in dynamic range in SFR afforded by our sample). While a thorough analysis of this low-redshift discrepancy is beyond the scope of this paper, part of the difference may lie in how the outflow rates were calculated from the CO line wings, as \citet{lutz20} included only the wings of the broad CO component while \citet{fluetsch19} included the entire broad component (i.e. including emission at systemic velocities that may not actually be part of the outflow).

\begin{figure}
\begin{centering}
\includegraphics[width=\columnwidth]{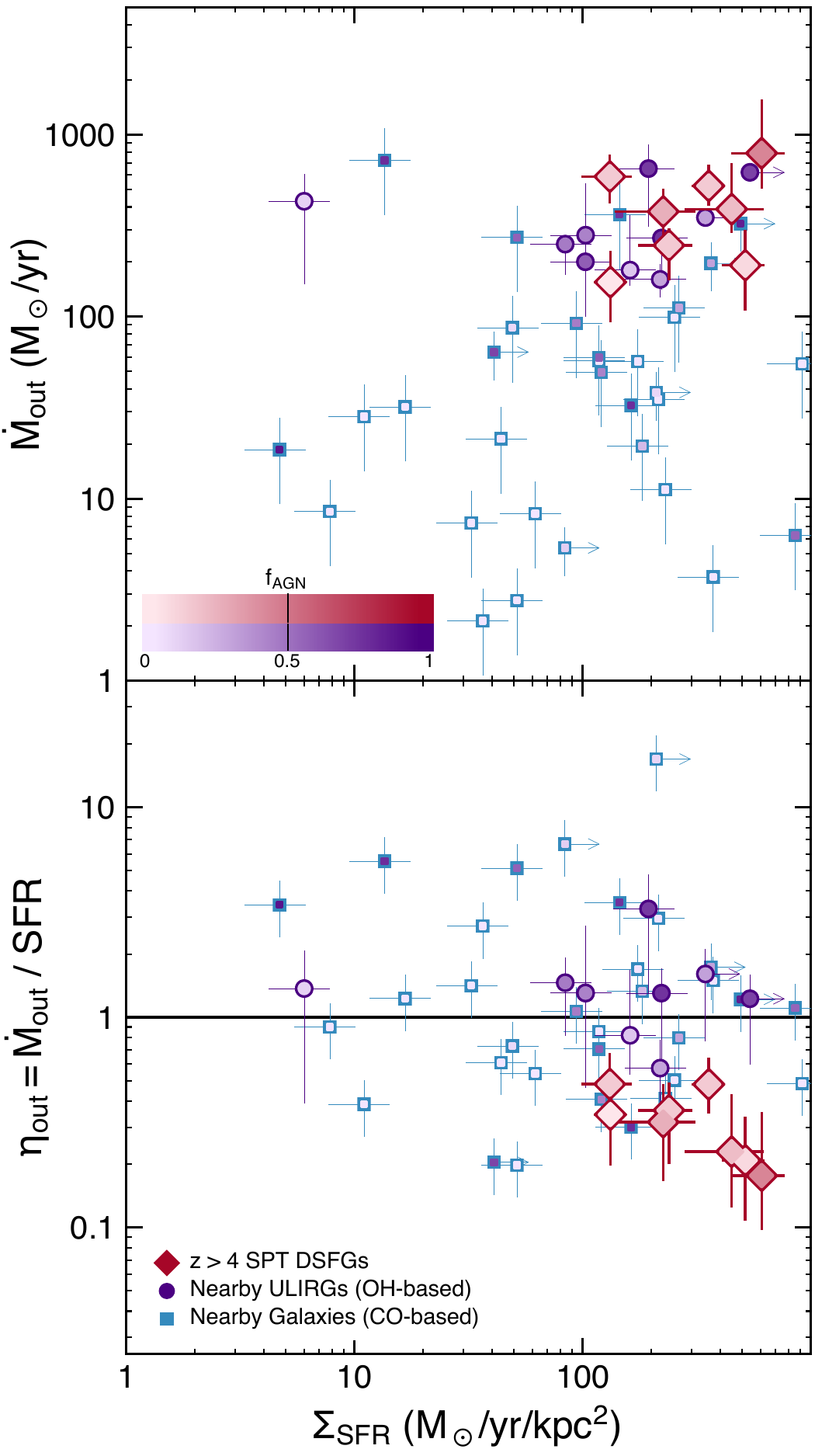}\\
\end{centering}
\caption{
As Fig.~\ref{fig:mdotoutsfr}, but as a function of the SFR surface density \sigsfr.
}\label{fig:mdotoutsigsfr}
\end{figure}

Figure~\ref{fig:mdotoutsigsfr} shows \Mdot and \etaout as a function of \sigsfr instead. The SPT sample lies well within the scatter but shows typically lower values of \etaout at a given \sigsfr compared to the low-redshift samples. As noted above, this is at least partially explained by the overall sub-linear trend between \Mdot and SFR. Similar to our investigation of outflow velocities with \sigir above, we again find no significant correlation between these properties, a conclusion that would not change if we increased the CO-based outflow rates by 0.5\,dex or adopted 2$\times$ higher limits on \fagn for our sample. As noted by \citet{lutz20}, however, the combined literature sample (and our own, clearly) is not complete in \sigsfr. Severe selection effects stemming from the diverse selection criteria in individual studies comprising the combined literature sample may exist, particularly at low \sigsfr.

\begin{figure}
\begin{centering}
\includegraphics[width=\columnwidth]{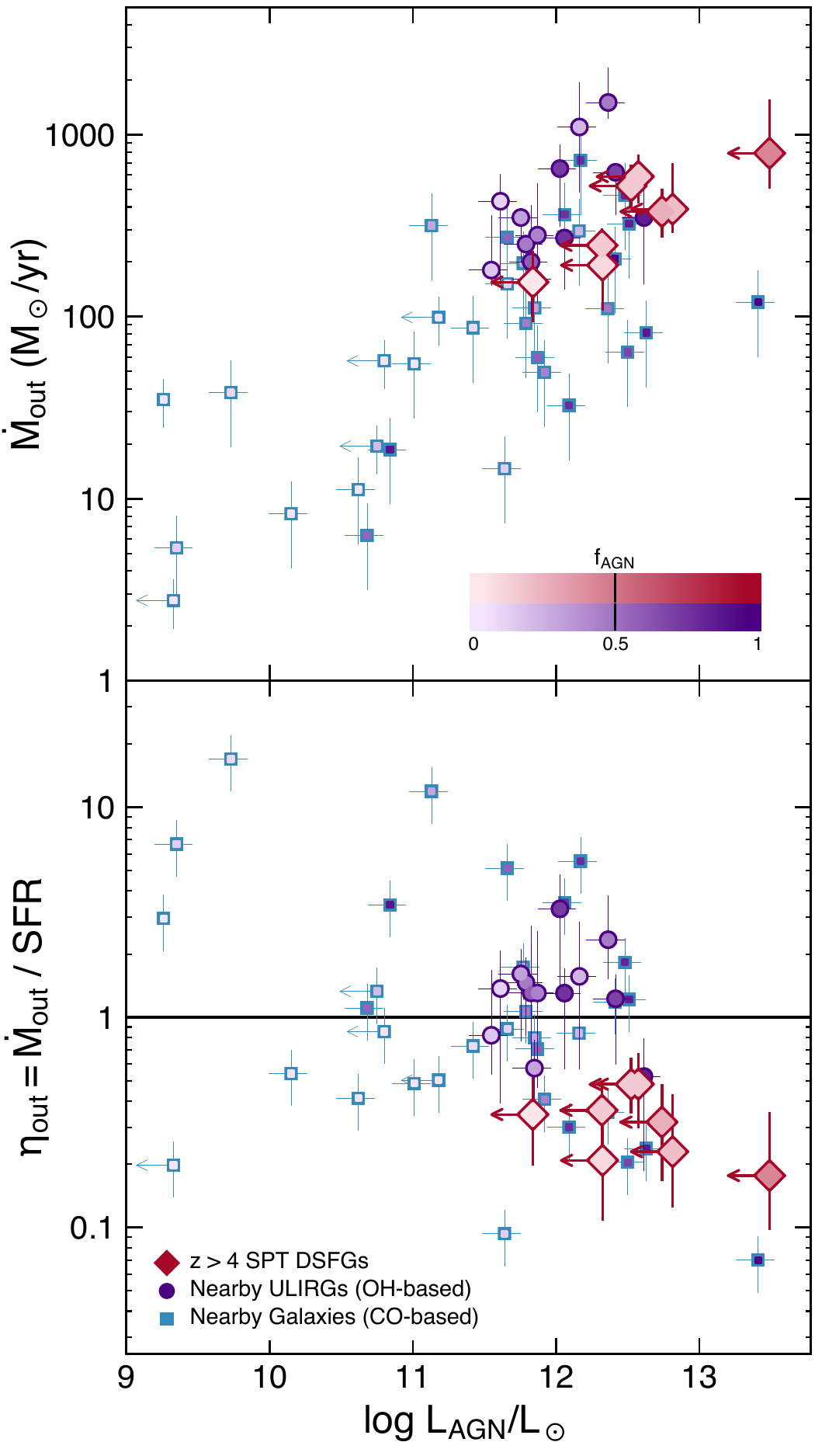}\\
\end{centering}
\caption{
As Fig.~\ref{fig:mdotoutsfr}, but as a function of the AGN luminosity \LAGN.
}\label{fig:mdotoutlagn}
\end{figure}

Finally, Figure~\ref{fig:mdotoutlagn} shows \Mdot and \etaout as a function of \LAGN. As seen in previous works, the low-redshift combined sample shows a clear relationship between \Mdot and \LAGN, with substantial scatter that increases at low \LAGN. The distribution of low-redshift objects in this parameter space has been discussed extensively in the literature \citep[e.g.][]{lutz20}. For our sample, we find that the limits on \LAGN from the available rest-frame mid-IR data do not result in the high-redshift objects being clear outliers in Fig.~\ref{fig:mdotoutlagn}, and they would not be outliers in the event that our limits on \fagn were underestimated by a factor of 2. As before in Section~\ref{driving} but from a different perspective, the outflows we have detected do not require an AGN based on mass loading factor scaling relations, but AGN activity also cannot be ruled out in our objects.

\subsection{Outflow Masses and Depletion Times} \label{outmass}

We now turn to estimates of the total molecular gas mass contained within the outflows. As described in Section~\ref{outflowgeom}, we use Eq.~\ref{eq:mdot} to calculate the outflow masses \Mout for our sample. For the low-redshift samples, we use the original published masses for both the OH-based and CO-based outflows. Recalculating the masses for the low-redshift samples using our assumed geometry results in $<$10\% differences in the median compared to the published values. Additionally, \citet{lutz20} find only a 0.06\,dex offset and 0.3\,dex dispersion between the masses for the low-redshift sources with outflows observed in both OH and CO. We make use of the total molecular gas masses for the low-redshift samples assembled by the original studies, all of which are based on low-$J$ transitions of CO (CO(3--2) or lower). The conversion factor between CO luminosity and \MHt is known to vary based on various galaxy properties \citep[e.g.][]{bolatto13}; we accept the values used by the original studies. For the SPT sample, we assume $\alphaco = 0.8$\,\Msol\,(K\,\kms\,pc$^2$)$^{-1}$, which we have previously found to be appropriate for the IR-luminous galaxies in our sample \citep[e.g.][]{spilker15,aravena16}.

\begin{figure}
\begin{centering}
\includegraphics[width=\columnwidth]{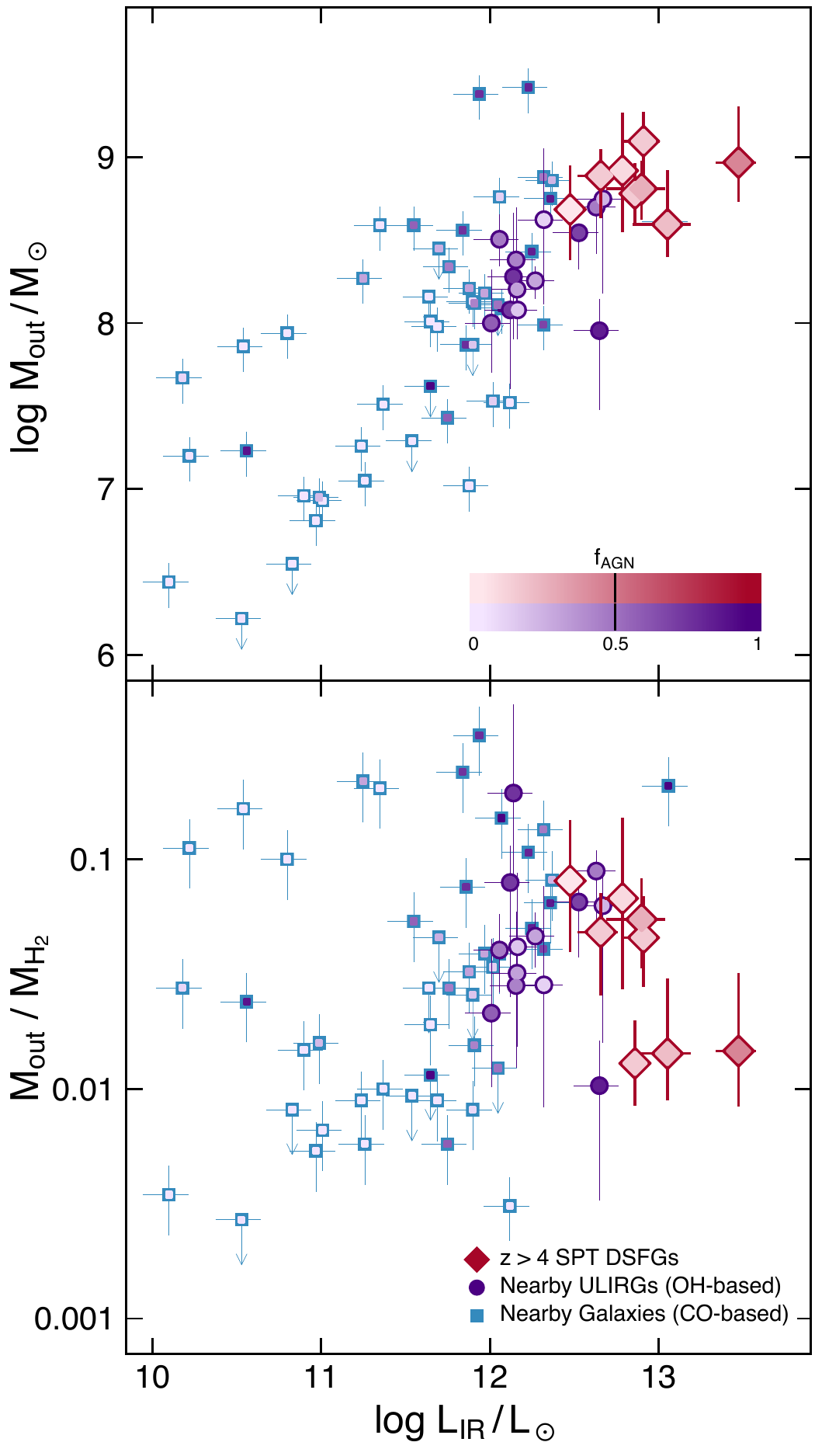}\\
\end{centering}
\caption{
Mass in the molecular phase of outflows (upper panel) and the fraction of the total galaxy molecular gas mass contained in the outflows (lower panel) as a function of \lir. Source symbols and color-coding as in Fig.~\ref{fig:mdotoutsfr}. 
}\label{fig:moutlir}
\end{figure}

Figure~\ref{fig:moutlir} shows the molecular outflow masses as a function of \lir, as well as the fraction of the total galaxy molecular gas mass contained in the outflows. Not unexpectedly, \Mout is clearly correlated with \lir, as has been previously noted many times in the literature. The $z>4$ SPT DSFG sample has molecular outflow masses in the range $\log \Mout/\Msol = 8.6-9.1$, unsurprisingly on the high end of the local samples. The masses of the two most intrinsically luminous sources in our sample, SPT2132-58 and SPT2311-54, are perhaps somewhat low in comparison to the extrapolation of the low-redshift samples, but are well within the observed scatter. From the lower panel of Fig.~\ref{fig:moutlir}, meanwhile, we find that the molecular outflows in our sample contain 1--10\% of the total molecular gas masses of the galaxies. These values are well within the range typically seen in low-redshift galaxies. Further, we find no discernible trend between $\Mout/\MHt$ and \lir despite the increase in dynamic range in \lir afforded by our sample.

\begin{figure}
\begin{centering}
\includegraphics[width=\columnwidth]{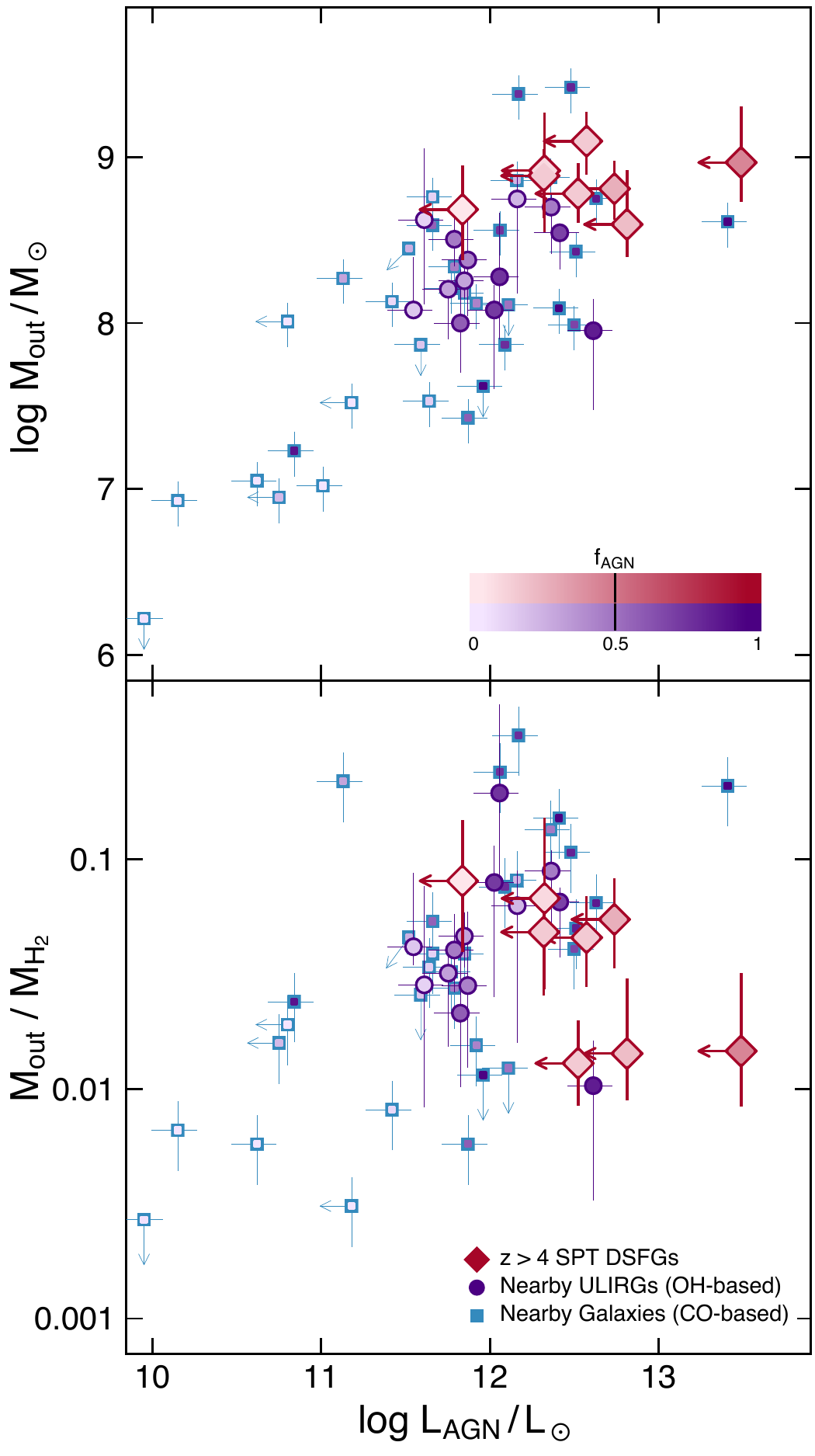}\\
\end{centering}
\caption{
As Fig.~\ref{fig:moutlir}, but as a function of the AGN luminosity \LAGN.
}\label{fig:moutlagn}
\end{figure}

Figure~\ref{fig:moutlagn} shows these same quantities as a function of \LAGN. Together, Figures~\ref{fig:moutlir} and \ref{fig:moutlagn} are effectively the corresponding versions of Figures~\ref{fig:mdotoutsfr} and \ref{fig:mdotoutlagn} for \Mout instead of \Mdot. As with the outflow rates previously, the current limits on \LAGN for the SPT DSFGs do not make them obvious outliers in Fig.~\ref{fig:moutlagn}. There are no indications from either of these figures that the dominance of the AGN plays any role either in determining \Mout in general or in defining the scatter in \Mout at a given \lir or \LAGN, as evidenced by the lack of secondary trends in these figures with \fagn.

\begin{figure}
\begin{centering}
\includegraphics[width=\columnwidth]{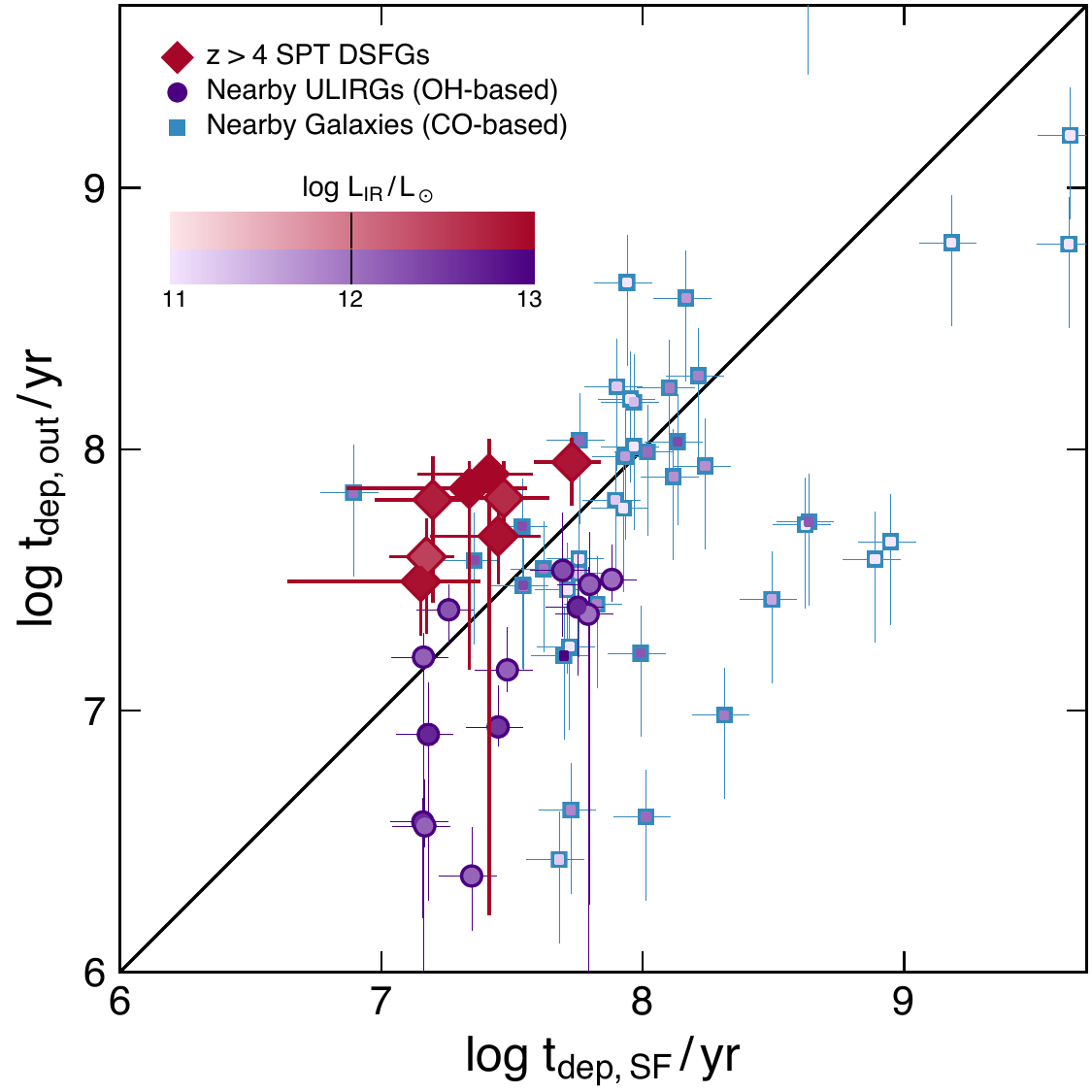}\\
\end{centering}
\caption{
Comparison of the molecular gas depletion timescales due to gas consumption through star formation ($\tdepsf \equiv \MHt$/SFR) and due to removal via molecular outflows ($\tdepout \equiv \MHt/\Mdot$). Symbols are as in Fig.~\ref{fig:mdotoutsfr}, but are now color-coded by $\log \lir$. The solid line indicates the one-to-one relation. Unlike the majority of low-redshift galaxies, we find shorter timescales for gas consumption by star formation than for removal in molecular outflows (a reflection of the sub-unity mass loading factors we find in our sample; Section~\ref{massloading}).
}\label{fig:tdepsfout}
\end{figure}

Meanwhile, Figure~\ref{fig:tdepsfout} compares the molecular gas depletion time scale due to outflows with that due to star formation. Here these depletion time scales are defined as the time it would take for the entire molecular gas reservoir of the galaxies to be removed by outflows or consumed by star formation, assuming the outflow rate or SFR remain constant. That is, $\tdepout \equiv \MHt/\Mdot$ and $\tdepsf \equiv \MHt$/SFR. These depletion times are only approximate estimates of the important time scales in the evolution of galaxies, given that both outflows and star formation operate simultaneously (which would give shorter depletion times), we ignore molecular gas destruction due to e.g. photo-heating or shocks (which would also shorten the depletion times), gas accretion and/or cooling into the molecular phase are neglected (which would give longer depletion times), and \Mdot and SFR are not in fact constant over time (which could push the depletion times either higher or lower depending on the time variability in \Mdot and SFR). Note that changing estimates of \MHt for any object in Fig.~\ref{fig:tdepsfout} moves objects diagonally parallel to the one-to-one line, since \MHt is incorporated in both axes.

For the $z>4$ sample, we find for all sources that $\tdepout \gtrsim \tdepsf$, a straightforward consequence of the sub-unity wind mass loading factors we determine (Section~\ref{massloading}). This conclusion would also hold if we artificially decrease the SFRs of our sample by doubling \fagn as a crude approximation of the effects of heavily-obscured AGN, though the depletion times would be about equal in that case. As before with \etaout, this places our sample with a distinct minority of the low-redshift samples. We stress again that all of these samples are biased towards galaxies and quasars that do host powerful outflows, and these results may not hold for objects with less extreme outflows that would be difficult to detect. Unsurprisingly given their very high SFRs and outflow rates, both depletion times are very short, $\sim$10--100\,Myr, and the fact that the two time scales are comparable points to the important role that outflows must play in regulating star formation in these galaxies.

\subsection{Outflow Momentum and Energetics} \label{outmom}

Galactic winds are often classified as either ``energy-driven'' if radiative losses in the outflowing gas are negligible or ``momentum-driven'' if they are not (regardless of the ultimate source(s) of the energy driving the wind). In the former case, the outflow is thought to be launched by the adiabatic expansion of a bubble of hot gas \citep[e.g.][]{chevalier85,silk98} that either lofts cold gas entrained in the expanding hot wind or (re-)forms cold gas from the swept-up shocked material at larger radii where the gas can radiatively cool \citep[e.g.][]{fauchergiguere12,costa14,richings18}. Similar to the energy-conserving Sedov-Taylor phase of supernova expansion, the resulting momentum in the outflowing gas can be `boosted' well above the radiative momentum flux driving the wind. In the momentum-driven case, in which radiative cooling is significant, momentum transferred to the gas from ram pressure or radiation pressure on dust grains results in gentler acceleration that may allow cold gas to reach large radii and high velocities before it is destroyed \citep[e.g.][]{murray05,murray11,thompson16,brennan18}. Real winds can of course be intermediate between these cases. For both momentum- and energy-driven winds and for winds driven by AGN or star formation, theoretical models provide estimates of the coupling efficiency between the input momentum and energy and the outflowing gas.

We calculate estimates of the outflow momentum and energy for our SPT sample as described in Section~\ref{outflowgeom}. For both the low-redshift OH- and CO-based samples we use the original published values of the outflow momentum and energy instead of those derived from our own assumptions in Sec.~\ref{outflowgeom}. A comparison between the published values and our estimates indicates that we may be overestimating the outflow momentum and energy by $\sim$30 and 70\%, respectively, while we find no systematic difference in the mass outflow rates. While still within the substantial uncertainties, this probably means that the values of \vef we use in Eq.~\ref{eq:momen} are higher than the outflow `characteristic' velocity, and some lower velocity between \vfifty and \vef would provide more accurate outflow energetics. We continue with our use of \vef; our conclusions here would be further strengthened if the outflow momentum and energy were systematically lower.

\begin{figure*}
\centering
\includegraphics[width=0.9\textwidth]{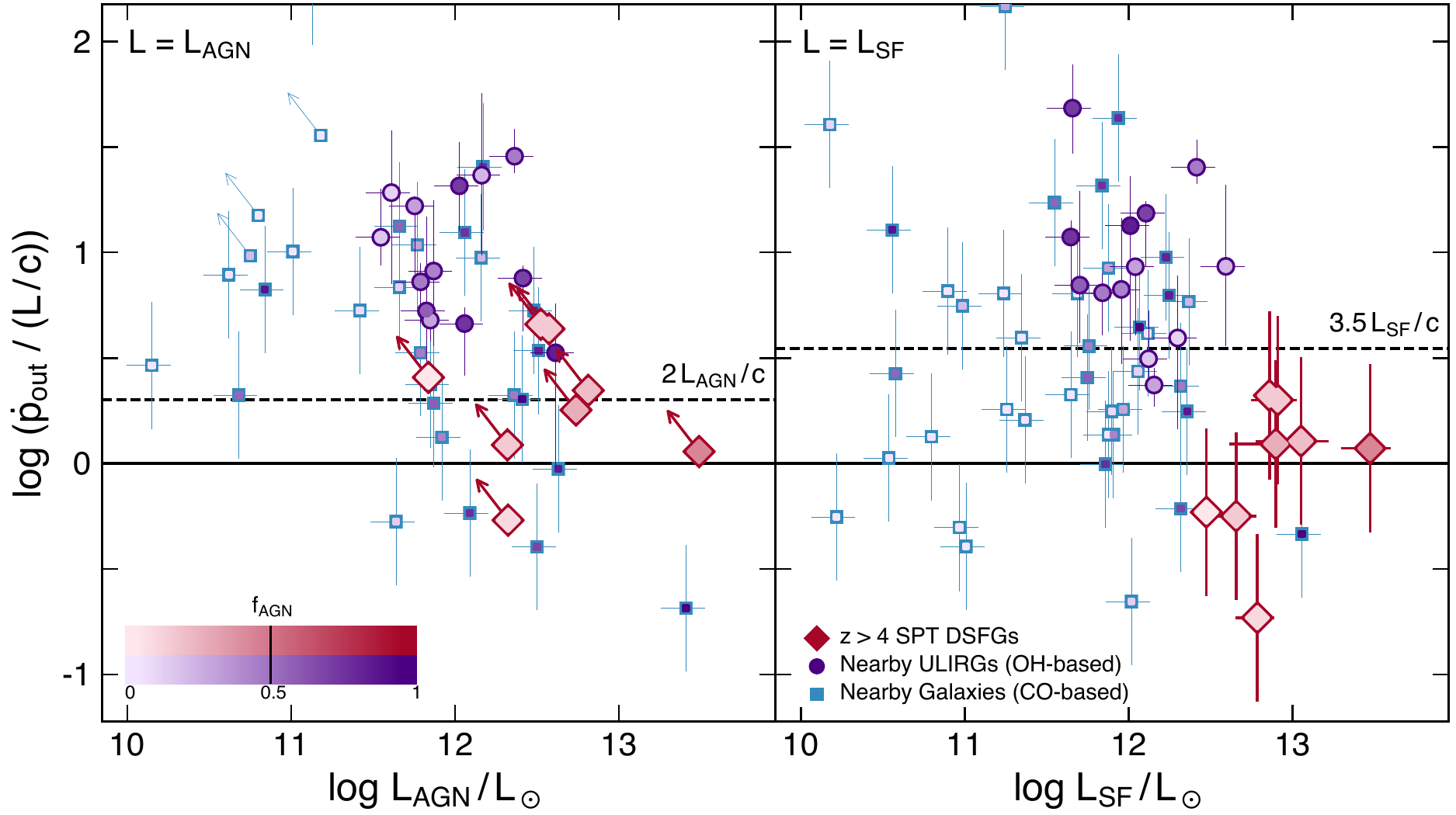}
\caption{
The ratio of the outflow momentum rate \pdot to the radiative momentum flux $\dot{p}_{\rm{rad}} = L/c$ provided by the AGN (left; $L = \LAGN$) or star formation (right; $L = \LSF$) as a function of the AGN and star formation luminosity. Symbols and color-coding as in Fig.~\ref{fig:mdotoutsfr}. Horizontal dashed lines indicate the approximate maximum momentum attributable to the AGN or star formation in momentum-driven winds. For the high-redshift SPT DSFGs, unlike in most local ULIRGs, the radiative momentum flux provided by star formation is fully sufficient to explain the observed outflows; neither AGN power nor energy-driven wind phases are required.
}\label{fig:pdotout}
\end{figure*}

In Figure~\ref{fig:pdotout} we show the fraction of the estimated outflow momentum rate compared to the total radiative momentum rate ($\dot{p}_{\rm{rad}} = L/c$) as a function of the estimated luminosities due to AGN and star formation for the sample of molecular outflows assembled from the literature and our high-redshift sample. In momentum-driven winds the AGN may provide a momentum rate up to $\sim$2$\LAGN/c$, treating both radiation pressure on dust grains and the AGN inner winds as $\LAGN/c$. Meanwhile, a continuous starburst can generate a maximum of $\sim$3.5$\LSF/c$ \citep{veilleux05,heckman15} through a combination of radiation pressure and the pressure of hot wind material driven by supernova ejecta. As seen in Fig.~\ref{fig:pdotout}, molecular outflows in low-redshift galaxies frequently show large momentum boosts $\sim$2--30 above the radiative momentum provided by the AGN and/or star formation, often taken as evidence that an energy-driven wind phase is required to achieve such large boosts (though see also \citealt{thompson15}, who argue that radiation pressure on dust grains can also achieve large momentum boosts in conditions possibly realized in very dusty and gas-rich galaxies). 

We find much more modest momentum ratios in our sample of high-redshift DSFGs, with maximum momentum boosts of $\sim$2 compared to the luminosity due to star formation, and all sources consistent with no momentum boost above the radiative momentum injection at all. This momentum boost is well within the range achievable by radiation pressure on dust in cases where the effective IR optical depth is of order unity \citep{murray05,thompson15}. Further, the momentum injection due to star formation alone is fully consistent with the observed outflow momentum fluxes; no additional radiative momentum from AGN is required. Indeed, it is not clear if the AGN alone could provide sufficient momentum to explain the observed outflows given the current limits on \LAGN; at least some substantial contribution from star formation would be required if AGN are relevant to the outflow energetics. This result would not change if we redistributed the total luminosity arising from the AGN and star formation by doubling \fagn compared to our current limits, although in that case the momentum flux from the AGN would also be sufficient to drive the outflows we observe. All sources would still show momentum boosts $\lesssim$3.5$\LSF/c$, would still be consistent with momentum-driving due to star formation, and would not show momentum boosts as large as those seen in the local ULIRGs. For our sources to be $>$1$\sigma$ inconsistent with the rough maximum $\sim$3.5$\LSF/c$ would require $\fagn > 0.8-0.99$ depending on the source, far above our current limits from the rest-frame mid-IR. Sources with such high \fagn typically show OH solely in emission in nearby objects (see discussion in \citealt{stone16}), while none of our sources show OH in emission. This could be taken as evidence that no source in our sample has $\fagn \gtrsim 0.9$.

\begin{figure*}
\centering
\includegraphics[width=0.9\textwidth]{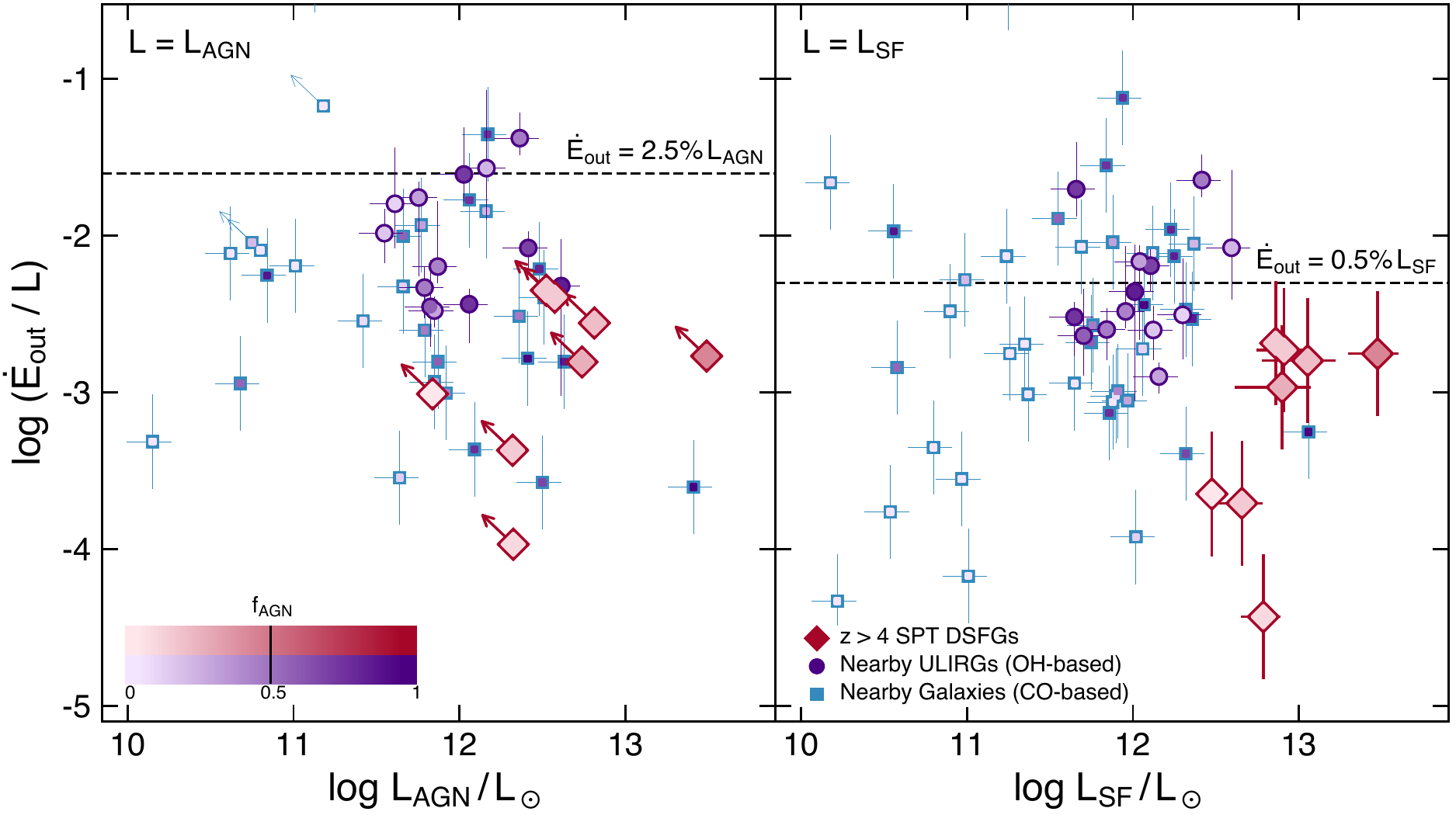}
\caption{
The ratio of the outflow kinetic power \Edot to the total luminosity of the AGN (left) or star formation (right) as a function of the AGN and star formation luminosity. Symbols and color-coding as in Fig.~\ref{fig:mdotoutsfr}. Horizontal dashed lines indicate the approximate maximum fractions of the AGN or star formation luminosity that can couple to the outflowing gas. For the high-redshift SPT DSFGs, unlike many low-redshift galaxies, the luminosity provided by AGN is not necessary to explain the outflow energetics; the outflows we have observed are fully consistent with the energy input from star formation alone.
}\label{fig:Edotout}
\end{figure*}

Figure~\ref{fig:Edotout} shows a similar plot for the outflowing kinetic power, following our outflow calculations in Section~\ref{outflowgeom}. Hot energy-driven winds from AGN are thought to be capable of supplying up to about $\sim$5\% of the AGN power to the outflows, of which some fraction $\sim$1/2 can plausibly be converted into bulk kinetic energy in the wind \citep[e.g.][]{king15,fauchergiguere12}. The mechanical luminosity generated by supernovae during a starburst, meanwhile, may reach $\sim$2\% of the total starburst luminosity, with perhaps $\sim$1/4 of this luminosity converted into kinetic motion in the ISM \citep[e.g.][]{veilleux05,harrison14}. The outflow energetics in many low-redshift molecular winds exceed the expected coupling efficiency to the starburst luminosity while the AGN energetics are in better agreement (Fig.~\ref{fig:Edotout}). This has been taken as evidence that the AGN must be primarily responsible for driving the low-redshift molecular outflows, and, in combination with the momentum rates in Fig.~\ref{fig:pdotout}, that these winds must be at least partially energy-driven.

In contrast to these low-redshift results, we find that the outflow kinetic energy rates in our $z>4$ DSFGs are uniformly below the threshold coupling efficiency for supernova-driven winds, and would still be consistent with this coupling efficiency if we adopt limits on \fagn twice as high (or more) as current data indicate. As with the momentum rates, AGN are not required in order to explain the observed outflow energetics. Moreover, the AGN in our sample could be an order of magnitude less luminous than the current limits without the outflow kinetic power approaching the theoretical maximum $\sim$few percent of the AGN luminosity.

Taken together, we conclude from Figures~\ref{fig:pdotout} and \ref{fig:Edotout} that (1) the high-redshift molecular outflows we have observed are fully consistent with expectations for momentum-driven winds, with no need for partially or fully energy-conserving phases, and (2) the observed outflow energetics can be fully explained by the momentum and energy provided by star formation alone in these galaxies, with no need for additional driving by AGN. We emphasize that we do not conclude that AGN are \textit{not} responsible for driving the observed outflows, merely that AGN are not \textit{required} to explain the energetics.  We note again that these conclusions are further strengthened if we adopt a somewhat lower characteristic velocity in Eq.~\ref{eq:momen} as it appears may be appropriate by comparison to the OH-based outflow energetics \citepalias{gonzalezalfonso17}. Similarly, our conclusions are also not changed in the event that our present limits on \fagn are underestimated by a factor of two (or more) due to AGN so heavily obscured they are not detectable in the mid-IR. In that case either the AGN or star formation could be the ultimate driving source, but the outflow energetics would still not require AGN momentum or energy injection or energy-conserving phases.

Both these results are surprising and counter to the conclusions typically reached in low-redshift studies. Conventional wisdom dictates that AGN are necessary to regulate galaxy growth in massive galaxies, in part due to scaling relations such as that between black hole mass and galaxy or bulge mass. Yet in our high-redshift rapidly star-forming galaxies, we find no need for AGN in order to explain the molecular outflow energetics we have measured. While our sample objects are more luminous than almost all of the low-redshift sources, we have no reason to expect that the outflow energetics should not also increase concomitantly with luminosity. Additionally, the energy-conserving wind mode is generally thought to have the highest coupling efficiency with the ISM, capable of sweeping up a large fraction of the gas in the ISM \citep[e.g.][]{zubovas12}. Our results, however, show that such high-efficiency energy-driven winds are not necessary to explain the observed outflow momenta and kinetic energy rates in our sample.

The differences between the $z>4$ DSFGs and nearby ULIRGs -- both with outflow properties from OH spectroscopy -- are particularly striking given their general similarities as highly dust-obscured and IR-luminous galaxies. \citetalias{gonzalezalfonso17} find that additional energy injection from AGN is required to explain the energetics of most of the low-redshift molecular outflows in ULIRGs (Fig.~\ref{fig:Edotout}) and that at least partially energy-conserving wind phases are likely necessary to explain the large momentum boosts (Fig.~\ref{fig:pdotout}). Neither of these appears to be true for the high-redshift DSFG outflows. We also note that a similar conclusion appears to be true for the only other $z>4$ object with detected OH absorption, a $z=6.1$ quasar where we expect the total luminosity to be dominated by the AGN \citepalias{herreracamus20} in contrast to our own sample with only upper limits on \LAGN. Although an estimate of the AGN and starburst luminosities separately is not available for this source and the OH detection was low-significance, applying the same outflow property calculations to this source as our sample would also place it in the general vicinity of our sample objects as long as $\fagn\gtrsim$0.1, a condition easily met for luminous quasars. 

It is tempting to ascribe at least some of the differences we see compared to the nearby ULIRGs to the overall difference in luminosities between the low- and high-redshift sources. Due to observational limitations the high-redshift objects are typically several times more luminous than the low-redshift ULIRGs. Increasing the luminosity of the ULIRGs would move them down and right in Figs.~\ref{fig:pdotout} and \ref{fig:Edotout}, in the direction that would be required to unify the low- and high-redshift objects. However, this would imply that the outflow momentum rates and kinetic power have essentially reached their maximum in low-redshift ULIRGs and no longer continue to increase in more-luminous systems as observed locally \citepalias{gonzalezalfonso17}.  It is also possible that the physics of outflows is qualitatively different between the low- and high-redshift samples. Multiple simulation efforts have found that star formation-driven outflows become inefficient in massive galaxies at $z\lesssim1$, so it could be that the low-redshift samples are predisposed towards AGN-driven winds by virtue of the fact that they have outflows detected at all \citep[e.g.][]{muratov15,hayward17}. It is clear that a larger sample at high redshift that spans a wider range in parameter space than current observations will be required to understand the dependencies of outflow energetics on galaxy properties.

There is also a probable selection effect that appears to be at play in the low-redshift sample. As shown in Figure~\ref{fig:kinematics}, while our sample overlaps with the low-redshift samples by most metrics, the subset of low-$z$ sources selected for detailed OH radiative transfer modeling by \citetalias{gonzalezalfonso17} have preferentially higher outflow velocities than the low-redshift sample overall, likely because these sources presented a more tractable sample for their modeling. This may weight the low-redshift sample towards AGN-driven (fast) outflows. Additionally, a bias towards fast winds can sharply skew the outflow energetics because the outflow velocity enters at least linearly in the outflow momentum rates and at least quadratically in the kinetic power (the outflow rates themselves are also proportional to \vout).  We thus expect that the local ULIRGs with slower outflows would show substantially lower momentum and kinetic energy outflow rates that extend down to the values we find for the SPT sources. For the majority of nearby ULIRGs, then, we expect that the outflow energetics would also be consistent with momentum-driven winds that do not require additional energy injection from the AGN.

\subsection{Fate of the Outflowing Gas} \label{escape}

The molecular outflows we have observed could plausibly affect the host galaxies over cosmological timescales, especially if large fractions of the cold gas in the outflows travel at sufficiently high velocity to escape the galaxy or even the dark matter halo virial radius. In the latter case, now unbound, the gas may never again be available for star formation. In the former, the gas becomes part of the circumgalactic medium and could recycle back into the galaxy unless continued energy injection or shock heating prevents the gas from cooling and condensing \citep[see][for a recent review]{tumlinson17}. 

We make a simple estimate of the fraction of the outflowing molecular gas that will escape the host galaxies by assuming the outflowing mass as a function of velocity is directly proportional to the equivalent width as a function of velocity, excluding the absorption components centered on systemic velocities \citepalias{spilker20}. To estimate the galaxy escape velocity for each source, we assume a spherical isothermal mass distribution truncated at a maximum radius $r_\mathrm{max}/r = 10$, following \citet{arribas14}. Because the detection of absorption requires the presence of continuum emission, we take $r$ to be the circularized effective size of the dust emission from our lensing reconstructions, \rdust. We estimate galaxy masses from total molecular gas masses based on CO(2--1) observations, assuming a typical gas fraction for DSFGs at these redshifts \citep{aravena16}. These masses are in reasonable agreement with simple dynamical mass estimates using the available \cii or CO line widths and the lens model sizes \citep[e.g.][]{spilker15}. We find escape velocities for our sources ranging from $\sim$400--1000\,\kms (median $\sim$700\,\kms), which agree reasonably well with other simple estimates scaling from the CO or \cii line widths or assuming pointlike mass distributions within \rdust. Given the uncertainties in mass and shape of the gravitational potential, we estimate typical uncertainties on the galaxy escape velocities of $\approx$40\%.

In this calculation, we ignore any additional deceleration of the outflow caused by sweeping up additional material. We have also implicitly assumed that the outflowing material is located at a typical distance from the galaxy center equal to the dust continuum emitting size, which seems reasonable based on our lensing reconstructions of the outflow material \citepalias{spilker20}, but we cannot rule out that much of this material is located deeper within the gravitational potential wells of the host galaxies. Both these effects would lower the fraction of outflowing gas that escapes the galaxies. On the other hand, we also assume that the outflowing gas is no longer being accelerated, which may not be the case if the winds are driven by the outward radiation pressure on dust grains, especially in the event of high far-IR optical depths and/or cosmic ray pressure. This would result in higher outflow escape fractions than our estimates.

\begin{figure*}
\centering
\includegraphics[width=0.65\textwidth]{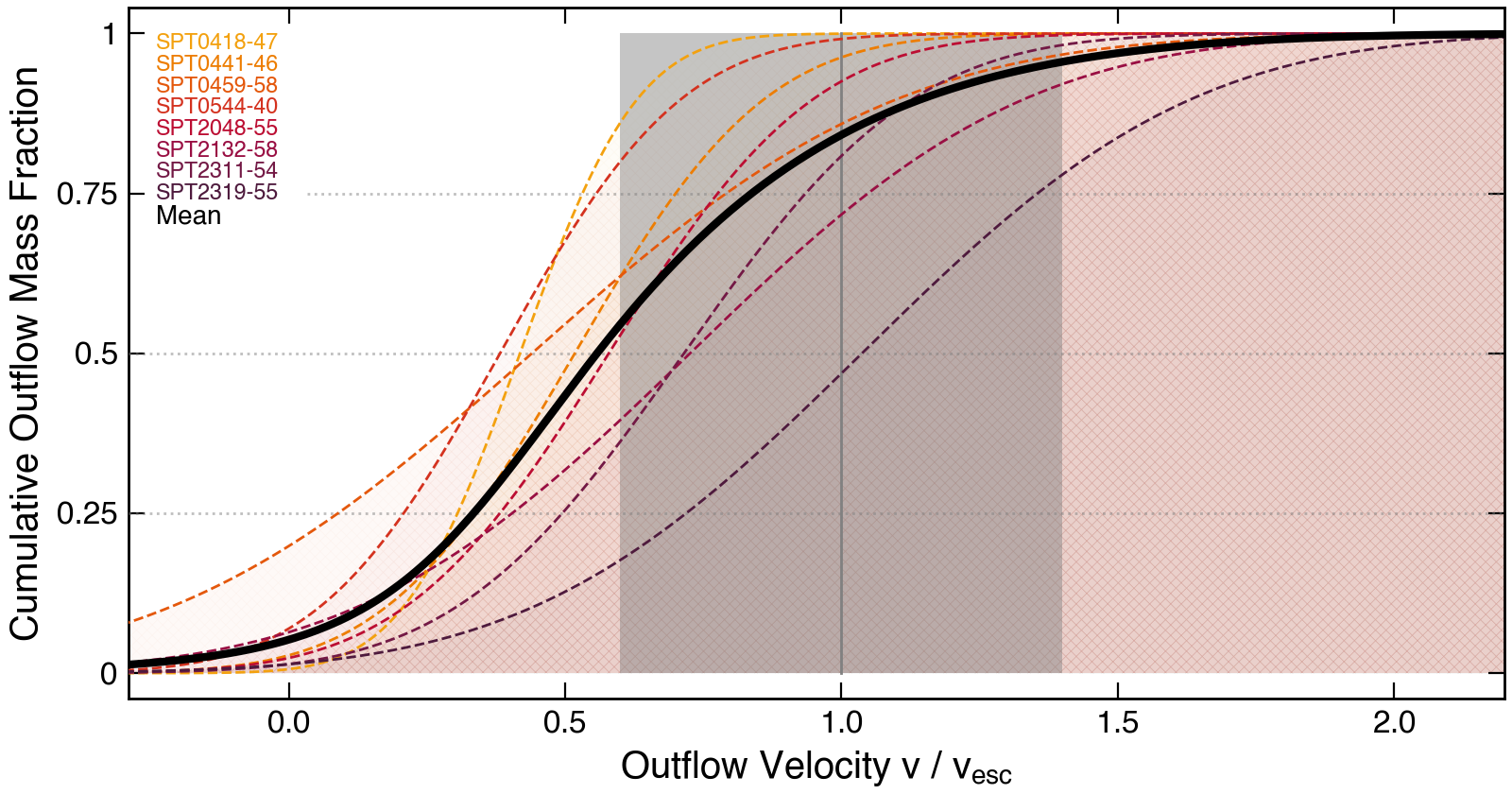}
\includegraphics[width=0.34\textwidth]{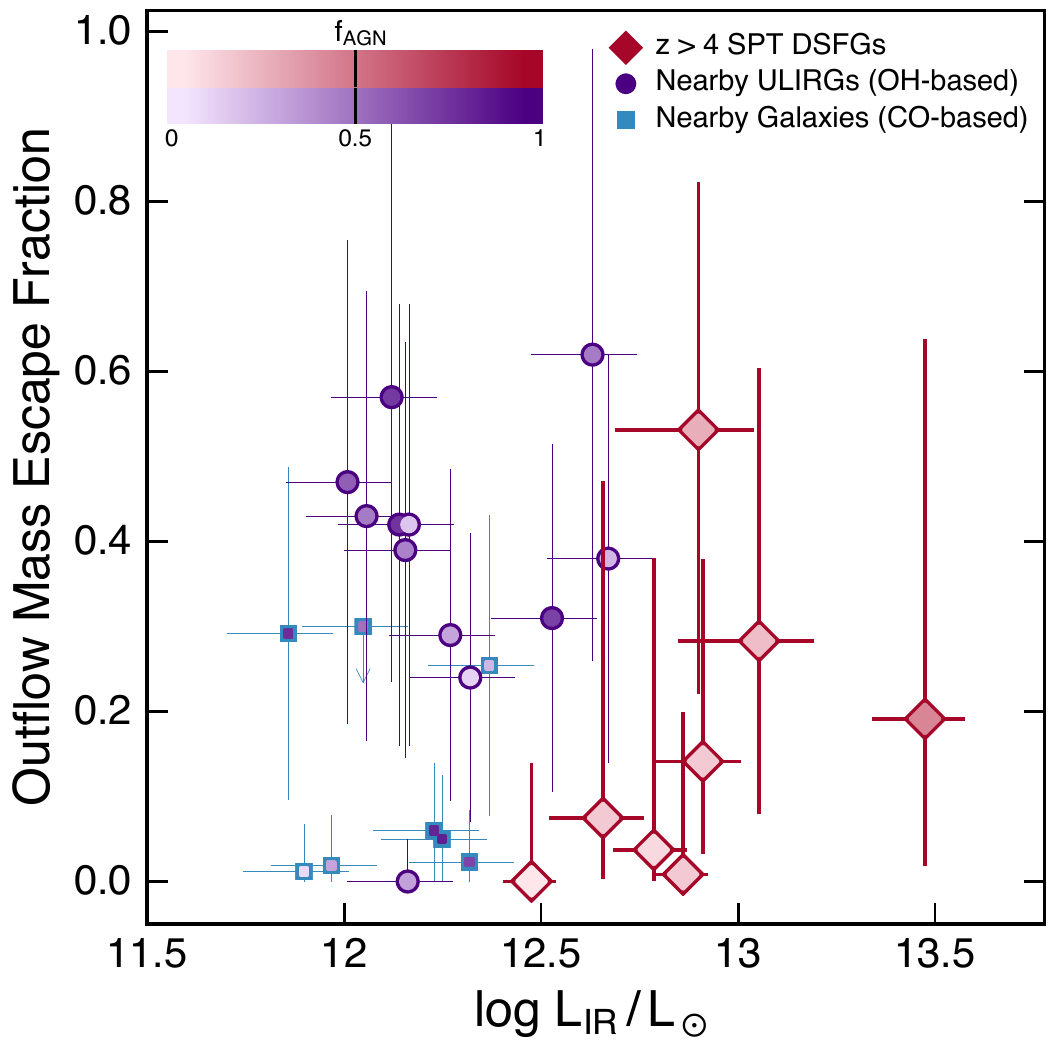}
\caption{
\textit{Left:} Cumulative molecular outflow masses as a function of the outflow velocity normalized to the galaxy escape velocity of each object; material moving at $v > v_{\mathrm{esc}}$ will leave the galaxy and enter the CGM. The thick black line shows the mean of the individual sample objects, while the vertical shaded region shows the approximate uncertainty on the escape velocities. 
\textit{Right:} Fraction of the outflowing material that will escape into the CGM as a function of \lir. Symbols and color-coding as in Fig.~\ref{fig:mdotoutsfr}. On average only $\approx$20\% of the molecular gas in the $z>4$ DSFG winds we have observed is traveling fast enough to leave the host galaxies, but there is wide dispersion in this fraction within the sample.
}\label{fig:vescgal}
\end{figure*}

Figure~\ref{fig:vescgal} shows the cumulative outflow mass for each object in our sample as a function of the outflow velocity, normalized to the estimated escape velocity. We find typical galaxy escape fractions $\sim$20\% with large variation within the sample. The three objects with estimated escape fractions $>$25\% are SPT2132-58 and SPT2311-54, which have the fastest outflows of our sample, and SPT2319-55, which has an atypically low mass given its outflow velocity (or an atypically fast outflow given its mass). Only $\lesssim$10\% of the outflowing gas is traveling at 1.5 times the escape velocity or faster, and essentially none is traveling at twice the escape velocity.

Figure~\ref{fig:vescgal} also shows these escape fractions as a function of \lir, now including the local galaxy samples with stellar velocity dispersion measurements available to estimate the escape velocities. The uncertainties on the escape fractions include only those due to the uncertain escape velocities, but do not include the (unknown) contribution from any variations in the equivalent width to outflow mass proportionality (or the CO--H$_2$ conversion factor for the CO-based masses). We find a very wide range in estimated galaxy escape fractions from $\approx$0 up to 60\%, reflective of the large range in outflow velocities and a rather limited dynamic range in galaxy mass. The escape fraction shows no obvious correlation with \lir or other observables, in agreement with our conclusions in Section~\ref{driving}.

The galaxy outflow escape fractions in Fig.~\ref{fig:vescgal} are substantially higher than those found by \citet{fluetsch19} even for the same objects. As discussed in Section~\ref{massloading}, this is because those authors count the full broad CO component as belonging to the molecular outflows, thereby including a substantial amount of CO flux at systemic velocities that need not actually be outflowing. This additional flux (and therefore mass) artificially lowers the galaxy escape fractions well below the values we obtain following the more conservative definition of \citet{lutz20}, who only consider the flux in the broad line wings in the outflow definition (excluding the core emission at systemic velocities). This more conservative definition results in total outflow masses a factor of $\approx5$ lower on average for the CO-based objects in Fig.~\ref{fig:vescgal} and consequently higher escape fractions compared to those found by \citet{fluetsch19}.

While the uncertainties are large, the nearby ULIRGs in Figure~\ref{fig:vescgal} tend to show somewhat larger escape fractions than our own sample of high-redshift objects. These sources have a mean and median escape fraction $\approx$40\%, about double that for our own sample. As previously discussed, this is most likely due to the fact that the sources with available OH-based radiative transfer models are preferentially also those with the fastest outflows  (Fig.~\ref{fig:kinematics}). Given the lack of correlation between outflow velocity and stellar velocity dispersion or stellar mass over the limited dynamic range probed by these samples \citep[e.g.][]{veilleux13}, this results in outflow escape fractions skewed towards larger values. As in Section~\ref{outmom}, we expect that a more complete sample of local ULIRGs would show substantially more overlap with the lower escape fractions we find for the high-redshift DSFGs.

The bulk of the molecular gas in the outflows is destined to remain within the galaxies, where it can become available for future star formation through a galactic fountain flow. At least in the cold molecular phase, most of the gas will not be permanently expelled and therefore these outflows cannot really be responsible for the very low gas fractions that are one of the hallmarks of quenched galaxies at lower redshifts \citep[e.g.][]{young11,davis16,spilker18b,bezanson19}.  Moreover, without continuous injection of thermal energy or turbulence over the long term, the CGM gas will develop a cooling flow resulting in significant gas accretion \citep[e.g.][]{su20}.

\subsection{Implications for Circumgalactic Medium Enrichment} \label{cgm}

\begin{figure}
\begin{centering}
\includegraphics[width=\columnwidth]{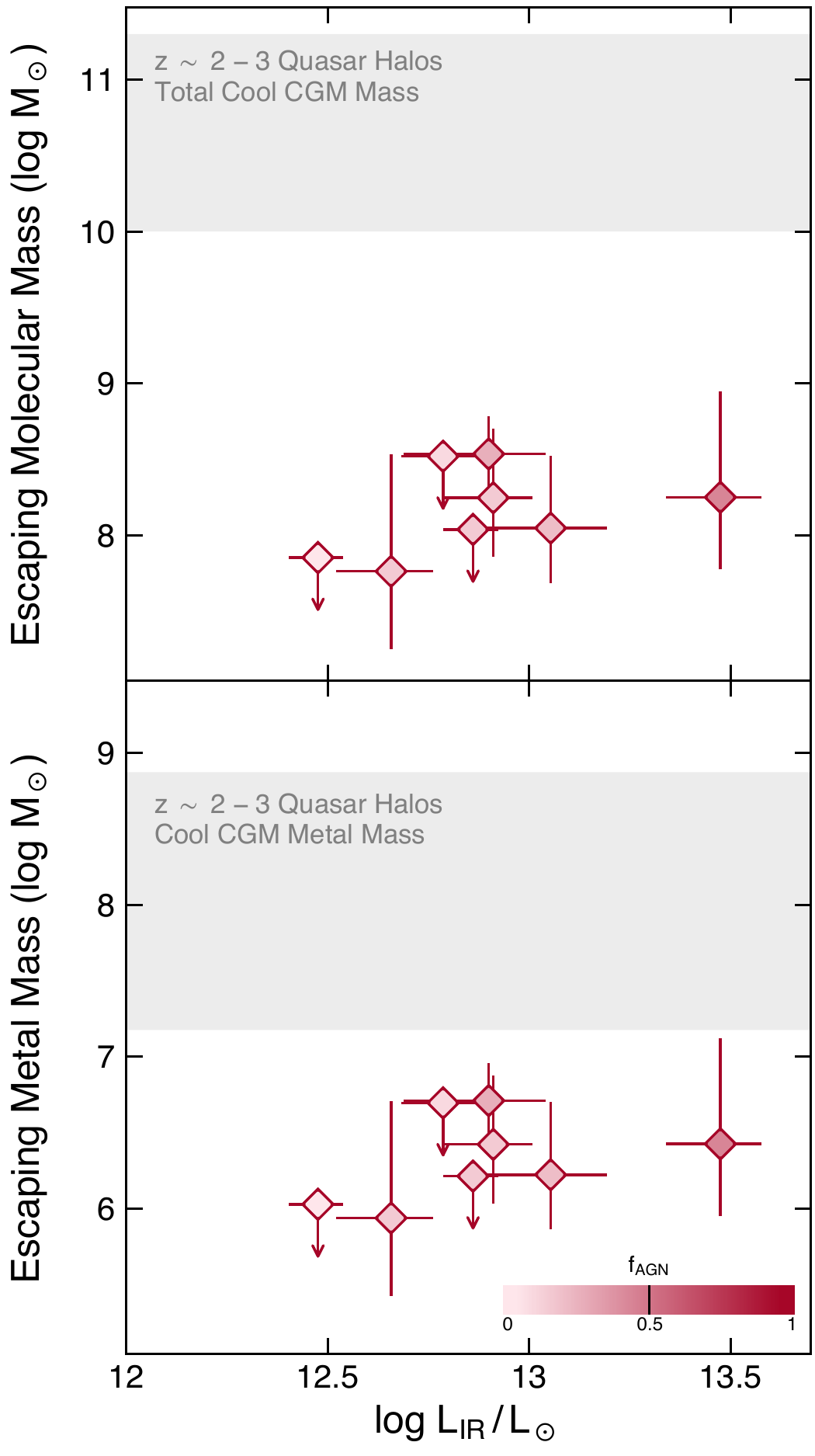}\\
\end{centering}
\caption{
Estimates of the molecular gas mass (upper panel) and metal mass (lower panel) contained in the observed $z>4$ SPT DSFG outflows that will escape the host galaxies and enter the surrounding CGM, assuming solar metallicity for the lower panel. For comparison, the grey shaded regions show the estimated total cool ($T \lesssim 10^4$\,K) CGM mass and metal mass surrounding quasar host galaxies at slightly lower redshifts \citep{prochaska14,lau16}. The molecular phase of the outflow episodes we have observed conceivably contribute $\sim$10\% of the total metals contained in the CGM at later times but only a small fraction of the total cool gas.
}\label{fig:cgmgas}
\end{figure}

Finally, we consider the impact of the outflowing molecular gas that probably \textit{will} escape in the context of the CGM surrounding these high-redshift DSFGs\footnote{We consider the outflowing gas to be entering the CGM if its velocity is greater than the galaxy escape velocity, but less than the halo escape velocity.}. The top panel of Figure~\ref{fig:cgmgas} shows the mass of the molecular outflows traveling at speeds greater than the galaxy escape velocity in each source. We assume all of this material enters the CGM, ignoring the loss of any material that escapes the larger dark matter halos (we expect this to be an exceedingly small fraction given the outflow velocity distributions in Fig.~\ref{fig:vescgal}). For most of the SPT DSFGs, we expect $\sim$few 10$^8$\,\Msol of the outflowing molecular gas to become incorporated into the CGM of the host halos.

The typical CGM properties of DSFGs are virtually unknown. Based on a sample of 3 $z\sim2$ DSFGs with background quasar sightline absorption spectra, \citet{fu16} speculate that the CGM of DSFGs may be less massive and/or that DSFGs inhabit somewhat less massive dark matter halos than co-eval quasars. However, given the much better statistics available for quasars at these redshifts, Fig.~\ref{fig:cgmgas} shows the typical range of total cool ($T \lesssim 10^4$\,K) CGM gas mass within the virial radius of $2 < z < 3$ quasar host galaxies thought to reside in $\log M_h/\Msol \sim 12-13$ mass halos \citep{prochaska14,lau16}.\footnote{Warmer CGM phases are extremely difficult to observe in the distant universe. Moreover, the thermal balance of CGM phases is an active area of investigation and subject to numerical resolution effects in simulations \citep{hummels19}. We restrict our analysis to the cool CGM phase for ease of observational comparison.} Given the possible differences between the CGM of DSFGs and quasars and an expectation that the CGM grows in mass from $z>4$ to 2.5, we expect this range to be an approximate upper bound on the total cool CGM mass surrounding the higher-redshift DSFGs in our sample.

The bottom panel of Fig.~\ref{fig:cgmgas} shows the total mass in metals being ejected into the CGM, under the simplifying assumption that the molecular outflows have approximately solar metallicity. If, as the outflow energetics suggest (Section~\ref{outmom}), processes related to star formation are responsible for driving the molecular outflows, we may expect the outflowing gas to be enriched significantly beyond solar, moving the points upwards in the lower panel of Fig.~\ref{fig:cgmgas}. In comparison, the metallicity of the cool CGM gas surrounding $2<z<3$ quasars is sub-solar, $Z \sim 0.1-0.3 Z_\odot$, likely because it is a mixture of metal-enriched outflow gas, less metal-rich material stripped or ejected from infalling satellites, and metal-poor material accreting from the cosmic web \citep[e.g.][]{muratov17,hafen19}. The range of total CGM metal mass for the same $z\sim2-3$ quasar samples is also shown in Fig.~\ref{fig:cgmgas}.

Taken together, the two panels of Figure~\ref{fig:cgmgas} give an intriguing (if admittedly speculative) picture of the relationship between molecular outflows and the CGM surrounding these galaxies. If the DSFGs in our sample will evolve to become like the quasars observed at slightly lower redshift, the current molecular outflow episodes will contribute only a small fraction $\lesssim$1--10\% of the total cool CGM mass. Evidently the total CGM mass must be assembled from some combination of outflowing gas in warmer phases than we have observed, many repeated outflow events through the lifetime of the galaxies, and accretion of additional gas into the CGM from the cosmic web or infalling satellites. While observations of the multi-phase components of outflows are rare even in the nearby universe, it appears that in general the molecular phase contains a significant if not dominant portion of the total outflow mass \citep{fluetsch19}, so additional mechanisms beyond accounting for the unobserved warmer phases are likely required. On the other hand, the current outflow episodes can contribute some substantial fraction $\sim$10\% or more of the total metals present in the CGM at lower redshift. This fraction would rise further if the outflowing molecular gas is enriched beyond solar metallicity. X-ray observations of the hot plasma in nearby winds typically find $\alpha$/Fe elemental abundance patterns (i.e. including oxygen, of relevance to our OH observations) enhanced to several times the solar value \citep[e.g.][]{nardini13,veilleux14,liu19}, though the composition of the molecular gas in outflows is unknown, even at low redshift.

The outflow metallicities of these highly-obscured galaxies are conceivably observable with future observations of far-IR fine structure lines \citep[e.g.][]{nagao11,pereirasantaella17}. Indeed, the \cii 158\,\um line has recently been detected on 10--30\,kpc spatial scales surrounding co-eval lower mass galaxies through stacking and, in a few cases, direct individual detections \citep{fujimoto19,fujimoto20,ginolfi20}. These studies conclude that metal-enriched outflows are the most likely source of the extended \cii emission, as generally expected from simulations \citep[e.g.][]{muratov15,hayward17,pizzati20}.

\section{Conclusions} \label{conclusions}

This work has focused primarily on deriving the physical properties of the largest sample of molecular outflows in the early universe to-date. These outflows, detected with ALMA as blueshifted absorption line wings in the ground-state OH 119\,\um doublet, appear ubiquitous among massive, IR-luminous DSFGs at $z>4$.  We rely heavily on observations of outflows in low-redshift galaxies with much richer OH spectroscopic data available, which we use as a `training set' of objects to derive outflow rates for our high-redshift sample with only the ground-state OH lines observed. Comparing four methods for estimating outflow rates, we find agreement at the factor-of-two level. Future improvements in the outflow rate estimates will require either observations of shorter-wavelength OH lines (e.g. the 79\,\um doublet) and/or the much less abundant $^{18}$OH isotopologue, both of which have far lower line opacities than the 119\,\um doublet currently available. Though the uncertainties on the outflow rates (and therefore the other outflow properties derived from the outflow rates) are large, we draw a number of conclusions from this first high-redshift outflow sample:

\begin{itemize}

\item We find tentative evidence that the outflow velocity correlates with \lir within the $z>4$ sample (Fig.~\ref{fig:kinematics} and Section~\ref{driving}). The same is not true for the combined low-redshift galaxies with OH data. A larger sample at high redshift will be necessary to determine whether there is a legitimate difference between outflows in low- and high-redshift objects.

\item We find high molecular outflow rates \Mdot ranging from $\sim$150--800\,\Msol/yr. This was not unexpected given the high IR luminosities of our sample. The wind mass loading factors are nevertheless slightly less than unity. The mass loading factors do not clearly correlate with any other quantity including SFR or \sigsfr. Gas consumption by star formation is more important than gas removal by outflows in regulating the molecular gas reservoirs of these objects (Figs.~\ref{fig:mdotoutsfr} and \ref{fig:tdepsfout}, Sections~\ref{massloading} and \ref{outmass}).

\item The cold molecular mass of the outflows is also high, $\log \Mout/\Msol \approx 8.5-9$. This still only represents 1--10\% of the total molecular gas mass of these gas-rich massive galaxies (Fig.~\ref{fig:moutlir} and Section~\ref{outmass}).

\item We find only very modest momentum boosts in the outflows compared to the radiative momentum, $\pdot / (L/c) < 3$. These boosts are fully achievable by winds driven either by supernovae or radiation pressure on dust grains. The outflow kinetic energy fluxes, similarly, are always less than the expected maximum values for outflows driven by star formation. There is no need for partially or fully energy-conserving wind phases (Figures~\ref{fig:pdotout} and \ref{fig:Edotout}, Section~\ref{outmom}).

\item Following the previous conclusion, the outflows we have observed do not require an additional injection of momentum or energy from AGN in these galaxies. While we currently have no evidence for AGN activity in our sample objects, with limits from rest-frame mid-IR photometry, we cannot rule out that deeply buried AGN are present. The outflow energetics, however, do not require AGN as the primary driving source.

\item We estimate that $\approx$20\% of the gas in the molecular outflows is traveling fast enough to escape the galaxies and enter the CGM, on average, though with large uncertainties and a range from 0--50\% within the sample. While an admittedly more speculative conclusion, we find that the molecular material moving fast enough to escape the galaxies represents only a small fraction of the total cool CGM mass but perhaps $\gtrsim$10\% of the metal mass observed in the CGM of massive halos at slightly lower redshifts (Figures~\ref{fig:vescgal} and \ref{fig:cgmgas}, Sections~\ref{escape} and \ref{cgm}).

\end{itemize}

While we have presented the largest currently-available sample of molecular outflows at $z>4$, it is by no means a cleanly selected or complete sample; our primary selection criterion was merely that the redshift of each target place the OH 119\,\um lines in an atmospheric window for ALMA observations. Given the high success rate in detecting outflows in these galaxies, we hope to have motivated future observations of samples that span a wider range in galaxy properties in order to build a more comprehensive view of the statistical properties of molecular outflows in the early universe. The physical properties derived for the outflows assembled from our present sample and future samples will provide invaluable constraints for simulations of galaxy evolution, tracking the prevalence and consequences of molecular outflows through the history of the universe.

\acknowledgements{
We thank the referee for a thorough and constructive report that improved the quality of this paper.
JSS is supported by NASA Hubble Fellowship grant \#HF2-51446  awarded  by  the  Space  Telescope  Science  Institute,  which  is  operated  by  the  Association  of  Universities  for  Research  in  Astronomy,  Inc.,  for  NASA,  under  contract  NAS5-26555. 
K.C.L., D.P.M., K.P., and J.D.V. acknowledge support from the US NSF under grants AST-1715213 and AST-1716127.
This work was performed in part at the Aspen Center for Physics, which is supported by National Science Foundation grant PHY-1607611.

This paper makes use of the following ALMA data: ADS/JAO.ALMA\#2015.1.00942.S, ADS/JAO.ALMA\#2016.1.00089.S, ADS/JAO.ALMA\#2018.1.00191.S, and ADS/JAO.ALMA\#2019.1.00253.S. ALMA is a partnership of ESO (representing its member states), NSF (USA) and NINS (Japan), together with NRC (Canada), MOST and ASIAA (Taiwan), and KASI (Republic of Korea), in cooperation with the Republic of Chile. The Joint ALMA Observatory is operated by ESO, AUI/NRAO and NAOJ. The National Radio Astronomy Observatory is a facility of the National Science Foundation operated under cooperative agreement by Associated Universities, Inc.

This research has made use of NASA's Astrophysics Data System.
}

\facility{ALMA}

\software{
CASA \citep{mcmullin07},
\texttt{visilens} \citep{spilker16},
\texttt{ripples} \citep{hezaveh16},
\texttt{astropy} \citep{astropy18},
\texttt{matplotlib} \citep{hunter07}}

\clearpage
\bibliographystyle{aasjournal}

\input{outflows_masses.bbl}
\end{CJK*}
\end{document}

%% file: table1_outflowrates.tex

\begin{deluxetable*}{lccccc@{\hspace{0.15in}}|@{\hspace{0.15in}}cccc} 
\tablecaption{SPT Sample Outflow Rate and Energetics Estimates \label{tab:sptmdots}}
\tablehead{
\colhead{Source} & 
\colhead{$\Mdot^{\mathrm{thin}}$} & 
\colhead{$\Mdot^{\mathrm{thin\,corr.}}$} & 
\colhead{$\Mdot^{\mathrm{S18}}$} & 
\colhead{$\Mdot^{\mathrm{HC20}}$} & 
\colhead{$\Mdot^{\mathrm{PLS}}$} & 
\colhead{$\Mdot^{\mathrm{joint}}$} & 
\colhead{$\Mout$} & 
\colhead{$\pdot$} & 
\colhead{$\Edot$} \\ 
\colhead{} & 
\colhead{$\Msol$\,yr$^{-1}$} & 
\colhead{$\Msol$\,yr$^{-1}$} & 
\colhead{$\Msol$\,yr$^{-1}$} & 
\colhead{$\Msol$\,yr$^{-1}$} & 
\colhead{$\Msol$\,yr$^{-1}$} & 
\colhead{$\Msol$\,yr$^{-1}$} & 
\colhead{$10^8\,\Msol$} & 
\colhead{$10^{35}$\,dyne} & 
\colhead{$10^8\,\Lsol$} 
} 
\startdata 
SPT0418-47 & $>$5 & 100$\,^{+100}_{-50}$ & 140$\,^{+190}_{-190}$ & 220$\,^{+180}_{-180}$ & 170$\,^{+60}_{-50}$ & 150$\,^{+80}_{-60}$ & 4.8$\,^{+4.1}_{-2.4}$ & 2.2 & 6.7 \\[0.05in] 
SPT0441-46 & $>$6 & 120$\,^{+110}_{-60}$ & 170$\,^{+190}_{-190}$ & 270$\,^{+180}_{-180}$ & 300$\,^{+170}_{-160}$ & 190$\,^{+110}_{-80}$ & 8.3$\,^{+10.3}_{-4.8}$ & 1.4 & 2.3 \\[0.05in] 
SPT0459-58 & $>$76 & 740$\,^{+500}_{-290}$ & 430$\,^{+190}_{-190}$ & 610$\,^{+240}_{-240}$ & 590$\,^{+200}_{-150}$ & 590$\,^{+190}_{-170}$ & 12.5$\,^{+6.3}_{-4.6}$ & 20.7 & 150 \\[0.05in] 
SPT0544-40 & $>$48 & 510$\,^{+350}_{-200}$ & 460$\,^{+230}_{-230}$ & 630$\,^{+260}_{-260}$ & 530$\,^{+160}_{-150}$ & 520$\,^{+160}_{-120}$ & 6.0$\,^{+3.2}_{-2.0}$ & 19.4 & 150 \\[0.05in] 
SPT2048-55 & $>$10 & 180$\,^{+160}_{-80}$ & 240$\,^{+180}_{-180}$ & 280$\,^{+180}_{-180}$ & 240$\,^{+130}_{-140}$ & 250$\,^{+60}_{-90}$ & 7.7$\,^{+3.5}_{-3.4}$ & 3.3 & 8.9 \\[0.05in] 
SPT2132-58 & $>$47 & 330$\,^{+200}_{-120}$ & 340$\,^{+240}_{-250}$ & 570$\,^{+320}_{-330}$ & 660$\,^{+320}_{-230}$ & 390$\,^{+310}_{-100}$ & 3.9$\,^{+4.4}_{-1.4}$ & 18.4 & 180 \\[0.05in] 
SPT2311-54 & $>$118 & 720$\,^{+440}_{-270}$ & 510$\,^{+200}_{-200}$ & 1290$\,^{+400}_{-400}$ & 1580$\,^{+780}_{-660}$ & 790$\,^{+770}_{-280}$ & 9.3$\,^{+11.0}_{-3.9}$ & 44.8 & 530 \\[0.05in] 
SPT2319-55 & $>$33 & 390$\,^{+280}_{-160}$ & 280$\,^{+180}_{-180}$ & 430$\,^{+210}_{-210}$ & 500$\,^{+210}_{-150}$ & 380$\,^{+130}_{-100}$ & 6.4$\,^{+3.1}_{-2.2}$ & 12.5 & 86 \\[0.05in] 
\enddata 
\tablecomments{Outflow rate estimates are described in the text as follows.
   $\Mdot^{\mathrm{thin}}$, $\Mdot^{\mathrm{thin\,corr.}}$: Section~\ref{optthin};
   $\Mdot^{\mathrm{S18}}$, $\Mdot^{\mathrm{HC20}}$: Section~\ref{simpempirical};
   $\Mdot^{\mathrm{PLS}}$: Section~\ref{plsempirical};
   $\Mdot^{\mathrm{joint}}$: Section~\ref{mdotcompare}.
   We use the joint estimates $\Mdot^{\mathrm{joint}}$ and associated uncertainties throughout
   the remainder of the text, subsequently dropping the `joint' superscript for simplicity.
   We estimate typical uncertainties on \pdot and \Edot of $\sim$0.4\,dex.
   This table is available in machine-readable format at \url{https://github.com/spt-smg/publicdata}.}
\end{deluxetable*}